\documentclass[review]{elsarticle}

\usepackage{lineno,hyperref}

\journal{Information and Software Technology}

\usepackage{graphicx}
\usepackage{enumerate}
\usepackage{enumitem}
\usepackage{amsmath}
\usepackage{stfloats}
\usepackage{xcolor}
\usepackage{soul}
\usepackage{amssymb}
\usepackage{subfigure}
\newcounter{qcounter}
\newcounter{acounter}
\newtheorem{theorem*}{Theorem}
\usepackage{color,soul}

\begin{document}

\begin{frontmatter}

\title{AFSCR: Annotation of Functional Satisfaction Conditions and their Reconciliation within i$^*$ models}


\author[CU]{Novarun Deb \corref{mycorrespondingauthor}}
\cortext[mycorrespondingauthor]{Corresponding author}
\ead{novarun@acm.org}

\author[CU]{Nabendu Chaki}
\ead{nabendu@ieee.org}


\address[CU]{Department of Computer Science and Engineering, University of Calcuttta, Technology Campus, JD-2, Sector III, Bidhannagar, Kolkata, West Bengal 700106, India}

\begin{abstract}
\noindent\textit{Context}: Researchers, both in industry and academia, are facing the challenge of leveraging the benefits of goal oriented requirements engineering (GORE) techniques to business compliance management. This requires analyzing goal models along with their semantics. However, most prominent goal modeling frameworks have no means of capturing the semantics of goals (except what is trivially conveyed by their nomenclature).

\noindent\textit{Objective}: In this paper, we propose the Annotation of Functional Satisfaction Conditions and their Reconciliation (AFSCR) framework for doing the same. The entire framework is presented with respect to i* modeling constructs. 

\noindent\textit{Method}: This is a semi-automated framework that requires analysts to annotate individual goals with their immediate goal satisfaction conditions. The AFSCR framework can then reconcile these satisfaction conditions for every goal and verify whether the derived set of cumulative satisfaction conditions is in harmony with the intended set of goal satisfaction conditions.

\noindent\textit{Result}: If the derived and intended sets of satisfaction conditions are in conflict, the framework raises entailment and/or consistency flags. Whenever a conflict is flagged, the framework also provides alternate solutions and possible workaround strategies to the analysts by refactoring the given i* model.

\noindent\textit{Conclusion}: In this paper we present a new framework that uses satisfaction conditions for going beyond the nomenclature and capturing the functional semantics of the goals within i* models. The analysis performed during the reconciliation process is generic enough and can be adapted to any goal modeling framework if required.

\end{abstract}


\end{frontmatter}


\section{Introduction}
Goal-oriented requirement models have become largely popular in the research community \cite{JennHork}. However, the levels of adoption of these frameworks in the industry have not been quite impressive. The research community has been actively trying to implement business compliance checking within goal models; but this requires capturing the semantics of goal modelling notations. This, in turn, motivates the need for developing sophisticated engineering techniques that allow better analysis and reasoning with i$^*$ models. 

Leveraging the benefits of i$^*$ models requires performing complex analytical processes like compliance management\cite{1-PSEER}, change management\cite{2-PSEER}, enterprise architecture maintenance\cite{3-PSEER}, and business life cycle monitoring\cite{4-PSEER}. These processes rely heavily on the \textit{semantics} associated with goals. Although methods exist for co-ordinating goals within and across actor boundaries of i$^*$ models, there are hardly any mechanisms for capturing the \textit{satisfaction conditions} of these goals. 

In this paper the authors have focused mainly on \textquotedblleft change management\textquotedblright as the running example. Whenever an organization decides to modify its designs due to evolving requirements, these changes are reflected in the underlying goal models. We have to reconcile a goal model every time it undergoes some change.  All the running examples shown in this paper involve change management. The satisfaction reconciliation operation is triggered whenever there is a change in the underlying goal model. This idea of keeping a running system aligned with its evolving business requirements has also been highlighted in \cite{Fuzzy}. The AFSCR framework, as proposed here, has a similar business environment setting and applies to i$^*$ models instead of the KAOS framework.


The main motivation of this paper is to propose a solution that tries to resolve satisfaction conflicts and suggests alternate solutions to enterprise architects that are semantically correct. The AFSCR framework demands enterprise architects to explicitly annotate goals with their \textit{goal satisfaction conditions} only.  The \textit{satisfaction conditions} of a goal describe what becomes true when a goal is achieved. These can be viewed as the post-conditions of a goal. In other words, the exercise of achieving the goal makes these conditions true. Later in the paper, we will describe the notion of \textit{cumulative satisfaction condition}. These can only be defined for a goal in the context of a goal model. They describe the refinement of the satisfaction conditions of a goal with the satisfaction conditions of direct descendant sub-goals in the goal model. Thus, while the \textit{satisfaction conditions} of a goal are \textit{context independent}, the \textit{cumulative satisfaction conditions} are \textit{context-dependent} since they can only be described for a goal in the context of the goal model within which it is situated. The annotation mechanism should be simple and easily implementable so that this additional task does not prove to be onerous for requirement analysts. The annotation language should neither be too formal, thereby reducing the accessibility of requirement analysts, nor too informal so that complex analysis and reasoning becomes difficult. 

The choice of controlled natural language (CNL)\cite{7-PSEER} is primarily because it lies somewhere in the middle of the spectrum of annotation languages. On the one end of this spectrum we have strictly formal languages that reduce the accessibility of requirement analysts, and on the other end of the spectrum we have natural languages that are so informal that it makes analysis and reasoning quite complex. Controlled Natural Language (CNL) seems to strike a perfect balance between these two extremes. Providing a repertoire of CNL sentence schemas for satisfaction annotation seems to be a popular solution for analysis and reasoning purposes. Formal annotations can be derived from CNL sentence schema instances. CNL is a peripheral matter and not central to the AFSCR framework as it works on formal annotations. CNL is just one of the suggested means for acquiring formal annotations. The machinery proposed in this paper can work on formal annotations derived using any other language as well. 

It should be mentioned here that the problem of reconciliation does not appear in softgoals and contribution links. Unlike hard goals, conflicts in non-functional requirements (or soft-goals) are identified and resolved in a completely different method. The NFR framework gives some insight into how such conflicts may be identified and resolved. Also, the predicates being derived from CNL schemas will have true or false values and softgoals do not have such hard satisfaction conditions. False predicates are represented using the negation operator $\sim$ and signify the non-satisfaction of some functional satisfaction conditions.

There has been a considerable amount of research on how stakeholders evaluate goal models and different types of evaluation frameworks and strategies have been proposed for such models \cite{R1-1, R1-2, R1-3}. In \cite{R1-1}, the authors compare the evaluation strategies of three different frameworks - \textit{jUCMNav, OpenOME,} and \textit{Tropos}. Only \textit{jUCMNav} and \textit{Tropos} support quantitative evaluation frameworks. The \textit{OpenOME} framework supports only qualitative satisfiability analysis. Sebastiani et al. have proposed \textsc{Goalsolve} and \textsc{Goalminsolve} in \cite{R1-2} that represents the entire goal model in CNF and uses SAT solvers to determine properties. The proposed framework works with qualitative satisfiability analysis of goal models using backpropagation. \cite{R1-3} elaborates on how both quantitative and qualitative evaluation of goal models can be done using the \textit{jUCMNav} framework. The AFSCR framework is also a qualitative satisfiability analysis framework that considers the satisfaction conditions of goals during backward propagation of satisfiability. However, due to changing business requirements, if the satisfaction conditions of high-level goals are changed, then forward propagation can also be deployed using the AFSCR framework to realign and reconcile the goal model.

\begin{figure}[t]
	\centering
	\includegraphics[width=0.9\textwidth, page=25]{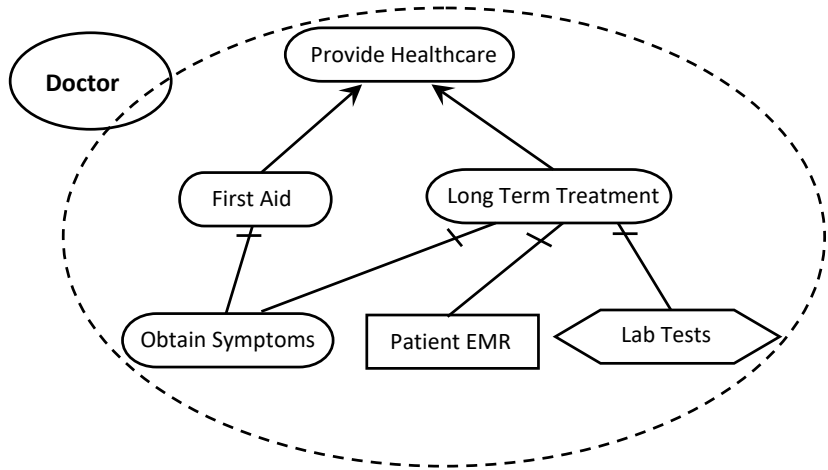}
	\caption{A simple i$^*$ model of an actor that shows two alternate strategies for \textit{Providing Healthcare}. \textit{First Aid} can be prescribed based on immediate patient \textit{Symptoms}. \textit{Long Term Treatment} requires the doctor to additionally look into the patient's past medical history (\textit{Patient EMR}) and may also require some \textit{Lab Tests} to be done.}
	\label{fig:ex_RHC}
\end{figure}

Consider the  i$^*$ model shown in Figure \ref{fig:ex_RHC}. The satisfaction conditions of the goals are apparently easy to understand from their nomenclature. However, if we look into the context of a particular patient, the satisfaction conditions associated with individual goals may give rise to conflicts. For instance, let us suppose that due to severe abdominal pain, a doctor suggests a CT scan of the abdominal area \textit{with contrast}. However, the patient's EMR may suggest that the patient is  \textit{allergic to contrast} fluids like iodine. So a patient's EMR must always be checked before prescribing any medical tests. This is not correctly captured by the i$^*$ model in figure \ref{fig:ex_RHC} and gives rise to a satisfaction conflict that is otherwise not detected by the nomenclature.
In this paper we try to identify and resolve two different types of satisfaction conflicts - \textit{entailment} and \textit{consistency} - in evolving business environments. We define what we mean by these two conflicts and give examples from the healthcare scenario.

\begin{list}{\noindent\textbf{Definition-\arabic{qcounter}}:~}{\usecounter{qcounter}}
	\item \texttt{Entailment conflict.} \textit{An i$^*$ model artifact $G_i$ within a goal model $G_M$ is said to have an entailment conflict if the cumulative conditions satisfied by the goal model subtree rooted at $G_i$ does not meet one or more of the immediate satisfaction conditions of $G_i$.}\\
	For example, w.r.t. figure \ref{fig:ex_RHC}, \textquotedblleft Lab Tests\textquotedblright needs to access \textquotedblleft Patient EMR\textquotedblright before performing the test; but the absence of resource \textquotedblleft Patient EMR\textquotedblright in the subtree rooted at \textquotedblleft Lab Test\textquotedblright (which is NULL in this case) gives rise to an entailment conflict.
	\item \texttt{Consistency conflict.}\textit{A pair of i$^*$ goals or tasks $\langle G_i, G_j\rangle$ will be deemed to be in a state of consistency conflict if $G_i \wedge G_j \models \perp$ (here we assume that the corresponding task or goal is written in the form of the truth-functional condition that represent post-conditions of executing a task or achieving a goal).}\\
	For example, w.r.t. the healthcare scenario, privacy regulations may impose permission requirements for accessing a patient's EMR. However, in case of emergency or accidents, a patient might not be conscious enough to grant access rights for treatment. This gives rise to inconsistent requirements which may prevent a doctor from providing proper healthcare.
\end{list} 

Let us assume that the goal model shown in figure \ref{fig:ex_RHC} a conflict-free goal model. Healthcare enterprises have to abide by certain medical rules. Let us assume that the medical standards board makes the following couple of changes in the medical regulations:

\begin{list}{\texttt{Change-\arabic{acounter}:}~}{\usecounter{acounter}}
	\item The healthcare enterprise passes a regulation that \textit{Long Term Treatment} cannot be provided without consulting a specialist.
	\item It must be ensured that patient is not allergic to any chemicals before performing a test. This is to prevent situations such as an MRI scan (with contrast) becomes the reason of death for a patient who is allergic to contrast fluids like iodine.
\end{list}
In the current state-of-the-art, goal modelers have to intervene manually and make necessary changes to the goal model so that these newly added regulations can be reflected in the enterprise. We will show later (in section \ref{sec:UseCase}) how the AFSCR framework can automate this process by only changing the annotated satisfaction conditions associated with the goals. 

The conditions governing correct decompositions of a goal provided by van Lamsweerde et al \cite{Lam98} are adequate for ensuring/checking the correct construction of goal models. This paper, however, addresses the question of how to address (and debug) common errors made by goal modelers. Also, in a changing business environment, a correct goal model specification may be rendered incorrect/erroneous due to changing business policies. Such a case study has been shown in Section \ref{sec:UseCase}. In both these situations, the AFSCR framework tries to automate the process of identifying and resolving the conflicts and generating a conflict-free goal model. This process helps in minimizing  errors at the later stages of the development life cycle. In the current scenario, goal model modifications have to be performed with the help of human intervention. Goal modelers have to look into the conflicts and resolve them manually. The AFSCR framework proposes a formal autonomous mechanism to aid goal modelers in the goal model modification process, thereby reducing the probability of introduction of human errors during the goal model modification process. 

The AFSCR framework consists of three different algorithms - the \textit{S}emantic \textit{R}econciliation \textit{A}lgorithm (\texttt{SRA}), the \textit{E}ntailment \textit{R}esolution \textit{A}lgorithm (\texttt{ERA}), and the \textit{C}onsistency \textit{R}esolution \textit{A}lgorithm (\texttt{CRA}). \texttt{SRA} reconciles the context-free \textit{satisfaction conditions} of goals into context-sensitive \textit{cumulative} satisfaction conditions. The \textit{cumulative} satisfaction condition annotations associated with any goal represents the semantics of satisfying the system requirements captured by the i$^*$ model \textit{sub}-tree rooted at that goal. Once the \textit{cumulative} satisfaction conditions of a goal are derived, \texttt{SRA} compares them with it's immediate satisfaction conditions (as annotated by the requirement analyst) and flags \textit{entailment} or \textit{consistency} issues, if they are detected. \texttt{ERA} and \texttt{CRA} are then used by the AFSCR framework to resolve entailment and consistency conflicts, respectively. Both these algorithms provide requirement analysts with possible \textquotedblleft$conflict-free$\textquotedblright alternatives, obtained by refactoring the original i$^*$ model.


The rest of the paper is organized as follows. Section \ref{sec:Review} provides a brief description of the works that have already been done on annotating i$^*$ models. Section \ref{sec:SEAnn} elaborates on how \texttt{SRA} works and raises flags on detecting conflicts. Section \ref{sec:ResCon} elaborates on the \texttt{ERA} and \texttt{CRA} algorithms and how they provide conflict-free alternatives to requirement analysts. A validation of the AFSCR framework is presented with the help of a real-world use case of the healthcare enterprise in section \ref{sec:UseCase}. Section \ref{sec:Analysis} presents an analysis of the properties of the algorithms introduced in the AFSCR framework. Finally, section \ref{sec:Concl} concludes the paper with a brief insight into the limitations of this work and possible future research directions.

\section{State-of-the-Art}
\label{sec:Review}
There has been very limited work in the existing literature that annotates i$^*$ models. This section highlights some of the research that has been done in the domain of annotating goal models with different types of attributes.

Liaskos and Mylopoulos \cite{ref1} have identified the sequence agnostic nature of standard goal modelling notations like i$^*$ \cite{Yu95} and annotated them with temporal logistics for deriving AI-based goal satisfaction planning. The authors introduce the notion of \textit{precedence links} and \textit{effect links} that annotate the i$^*$ model with preconditions and postconditions of fulfilling a goal. This kind of ordering allows formalization of goal models using temporal logics (like LTL, CTL, etc.). Although this method establishes some sort of a sequence between the tasks of a goal model, the notion of precedence does not remain intuitive for softgoals. Softgoal satisfaction can be facilitated with \textit{hurt} and \textit{help} contributions from tasks, hard goals, etc.

In \cite{ref2}, Liaskos \textit{et al.} have highlighted the importance of augmenting goal models (like i$^*$) with the optional requirements or preferences of the users. This paper uses the \textit{precedence} and \textit{effect} links proposed in \cite{ref1}. Additionally, this work introduces the notion of weighted contribution links for evaluating the degree of satisfaction or denial for softgoals. Accumulation and propagation of these weighted contributions follow the rules prescribed in \cite{ref2-9}. Optional user requirements are defined as \textit{Optional Condition Formulae} (OCFs) using first order satisfaction and domain predicates. Preferences are captured as linear combinations of OCFs and these preference formulae may be weighted or non-weighted in nature. Alternate goal plans are evaluated based on the degree of satisfaction of the preferences.

Koliadis and Ghose have been working with semantic effect annotations of business process models \cite{ref3,PSEER,PCTk}. In \cite{ref3}, the authors propose the GoalBPM methodology that maps business process models (using BPMN) to high-level stakeholder goals described using the KAOS framework \cite{Lam98}. This is done by defining two types of links - \textit{traceability} and \textit{satisfaction}. The former links goals to activities while the latter links goals to entire business processes. Satisfaction links require effect annotation of the business process model, followed by identification of a set of critical trajectories and, finally, identifying the subset of traceability links that represent the satisfaction links. In \cite{PSEER,PCTk}, the authors have worked with semantic effect annotation and accumulation over business process models (Process SEER) and how it can be extended to check for business process compliance using the PCTk toolkit.

Kaiya et \textit{al.}\cite{AGORA} have proposed the popular Attributed Goal-Oriented Requirements Analysis (AGORA) method that derives a goal graph from goal models and annotates the nodes and edges of the graph with attribute values and quality matrices. Attribute values consist of contribution values and preference matrices. Contribution values are used to annotate the edges of the goal graph and represent the contribution of the sub-goal towards the fulfillment of the higher-level goal. Preference matrices are the vertex annotations and represent the preference of respective goals to the concerned stakeholders. Both these attribute annotations can be used by analysts to choose among multiple alternate strategies, to perform conflict management and change management. Quality matrices are used to analyze the quality of the requirement specifications that are derived from the goal graphs. The metrics for measuring such quality may be correctness, unambiguity, completeness, inconsistency, etc. Yamamoto and Saeki \cite{ref7} have extended the idea of using annotated goal graphs for requirements analysis to software component selection.

The existing literature also documents works on consistency management of requirements other than GORE. There has been two directions of research for validation and consistency management of software requirements - heuristic analysis and formal analysis \cite{KMRD}. Heuristic analysis uses heuristic algorithms. For instance, in \cite{KMRD-21}, the authors propose a consistency checking tool called VIEWINTEGRA based on UML model transformations. Different types of UML models are transformed into consistent interpretations which are then compared for consistency checking.

Chitchyan et al. have proposed an automated extension of the RDL natural language processing tool suite that derives semantic annotations from requirement specifications \cite{KMRD-39}. It uses the MRAT Eclipse plug-in \cite{KMRD-40} to identify temporal dependencies within RDL compositions. The underlying hypothesis is that these temporal dependencies are the origins of sequence conflicts and inconsistencies.

The work of Kroha et al. \cite{KMRD-41} explores the possibility of identifying inconsistencies within requirement specifications with the use of semantic web ontologies. Ontological reasoning can identify contradictions but they cannot detect behavioral inconsistencies as they work with only the static parts of UML models, thereby ignoring the dynamic aspects.

\textquotedblleft Archetest\textquotedblright is a dual layered modeling environment that automates the process of identifying inconsistencies \cite{KMRD-44}. The authors claim that a wizard-based interface allows developers to create use-case descriptions more accurately. Although the initial results presented in this paper are interesting, they can be misleading as the framework needs to be tested on several different case studies. 

In \cite{KMRD-57}, the authors propose a consistency management scheme that uses a model composition scheme to create a global model from multiple heterogeneous models with the help of model composition. This requires mapping the heterogeneous models to an intermediate notation and then fusing them together. This approach allows the detection of inconsistencies. Traceability links between the original models and the global model helps in backtracking and identification of such inconsistencies.

Mehner et al. propose a more formal approach to consistency management in use case diagrams, activity diagrams and class diagrams of UML models \cite{KMRD-58}. Every activity is annotated with their pre- and post-conditions using a UML variant of that activity. A formal graph transformation technique AGG is used to annotate the UML variants with formal semantics, thereby allowing a formal analysis of consistency within the requirements. The proposed framework also analyses the interaction between functional and non-functional requirements.

\cite{KMRD-59} documents anther GORE based consistency management framework that uses the GOPCSD tool to check for inconsistencies and incompleteness within requirements. The tool adapts the goal-driven analysis technique from KAOS with the help of an animation interface that accounts the measures take for the aspect-based process control systems. The tool also has the added feature of transforming abstract user needs to formal requirement specifications followed by a transformation to B specification. The underlying assumption is that the requirements need to be corrected and validated by the user before the transformation can take place. Yu \cite{KMRD-61} also suggests an aspect-oriented consistency management scheme but for web applications only. The HILA tool uses extended UML state machines to model adaptation rules for web applications. However, the application domain can be extended to more generic situations thereby ensuring automated consistency checking for aspects and rules. 


There has also been a considerable amount of research in the direction of formal analysis for consistency management of requirements. In \cite{KMRD-18}, the authors develop a prototype called CARL which is capable of doing a brute-force search of use-case scenarios that can give rise to latent inconsistencies. This formal methodology begins with a preprocessing that uses the CARET reasoning engine to derive logical statements from natural language requirement specifications. CARET uses the Cicois syntax-based parser to parse requirement specifications documented in English. The tool combines heuristic optimization strategies and backtracking techniques to analyze inconsistencies in natural language specifications. The authors extend this tool by implementing theorem proving and model checking techniques in \cite{KMRD-67}. This extended tool allows the detection and handling of inconsistencies in a more formal manner. However, even this extension remains limited in its ability to model complex requirements due to the use of propositional logic.  

Nentwich et al. propose a inconsistency repair framework for  distributed requirement documents in \cite{KMRD-24}. The framework is capable of suggesting multiple interactive repair actions for each detected inconsistency. The limitation of this work lies in its inability to handle situations where the repair actions interact with the grammar of the requirement documents or with other repair actions generated from other constraints. The same research group has also proposed another lightweight consistency management framework called Xlinkit that uses first order logic \cite{KMRD-65}. The framework uses extended semantics and an incremental approach to produce hyperlinks that identify inconsistencies across different levels of specification. The framework reduces checking time but is unable to identify the problem associated with an inconsistency.  

The SC-CHECK tool \cite{KMRD-66} developed by Chen and Ghose is a consistency management tool that identifies inconsistencies in distributed requirements documents and provides industry-standard resolutions using semantic webs. The tool uses informal requirement specifications and a corresponding semi-formal representation to derive an abstract formal representation. Consistency rules are implemented using Prolog. The tool also provides an interactive interface to resolve the consistency violations.

Formal analysis of consistency management in GORE frameworks have been proposed by Lamswerde et al. in \cite{KMRD-70}. The authors propose KAOS as the requirements specification language  and both formal and heuristic techniques to identify violations of boundary conditions and domain constraints. Model checking is used to identify deviations from assertions and help in the elaboration of requirement concepts.

In \cite{KMRD-71}, Mu et al. propose a Viewpoints based framework for prioritizing requirements into high, medium and low categories. These prioritized views are then merged into a layered knowledge base. First order logic and prioritization values are used for trade-off analysis. Merging of viewpoints can, however, lead to unaccounted formulas representing merged viewpoint demands.

Scheffczyk et al. suggest domain specific resolution of inconsistent formal temporal constraints for functional requirements in business environments \cite{KMRD-73}. The CDET toolkit is a semi-formal tool for consistency management that allows the quality refinement of business requirement specifications. CDET manages consistency of requirements at different levels of abstraction and integrates them with industry processes using a revision control system (RCS). Although additional expertise is required for formalization of consistency rules, the tool is suited to work with computable properties in heterogeneous documents.

In \cite{KMRD-74} the authors check for inconsistencies within formal specifications which, in turn, are derived from CNL specifications. They use the B specification method to deploy first order logic for checking the inconsistency of requirements against constraints in safety-critical systems. Although the proposed method satisfies the safety property, the correctness and timeliness properties are not ensured. The proposed framework is also unable to support automated recovery from inconsistencies and incapable of handling complex business environments.

\textit{Techne} \cite{TKN} is a abstract requirements modelling metamodel that allows requirement engineers to build new requirement modelling languages for the early RE process. \textit{Techne} is derived from a schema of requirement specifications and uses atomic propositions in natural language for analysis. However, this framework is unable to support task sequencing and specification of temporal constraints. \textit{Techne} relies on knowledge representation for the identification of candidate solutions while handling inconsistencies. This demands the requirement engineers to identify conflicting goals and represent this knowledge in the model, thereby making semantic analysis a human process.

From the brief existing literature on semantic annotation of i$^*$ models, we see that researchers have attempted to annotate goal models with temporal information for simulation and model checking purposes. Researchers have also tried to identify satisfaction conditions of goal fulfillment for evaluating user preferences. In this perspective, it becomes necessary to spell out a mechanism for semantic  annotation of goals, and how these satisfaction conditions can be reconciled over the entire enterprise for performing different kinds of analysis. The literature survey on consistency management of requirements shows that, although GORE frameworks like i$^*$ can be very beneficial for business enterprises, there has been very limited research for consistency management in GORE frameworks, typically none using i$^*$. This paper tries to address this domain of consistency management in requirement specifications and propose an automated analyzer for consistency management in i$^*$ frameworks. \textit{Techne} has been proposed for generic goal models whereas AFSCR has been developed on the i$^*$ framework.\textit{Techne} is a very useful framework but it solves the problem of inconsistencies arising out of new goals being added in a goal model. The AFSCR framework addresses a different problem where requirement analysts may want to rationalize existing goal models. The paper on \textit{Techne} explicitly states that \textquotedblleft \textit{Techne and i$^*$ differ in several respects. i$*$ has no notion of	conflict, preference or mandatory/optional requirements, no formal semantics, and thus has no precise notion of what a candidate solution to the requirements problem is}.\textquotedblright  The AFSCR framework addresses all these drawbacks.
 
\section{Satisfaction Reconciliation}
\label{sec:SEAnn}
The main objective of this paper is to ensure consistency in the specification of goals and, thus, consistency in the shared understanding of organizational intent amongst stakeholders. We define an annotated i$^*$ model as one in which every model artefact has been annotated with their intended satisfaction conditions. These annotations can be represented as 2-pairs of the form $\langle$\textit{immediate-func, cumulative-func}$\rangle$. The term \textquoteleft \textit{func}\textquoteright in the above pair refers to functional satisfaction conditions as we intend to propose the framework with respect to functional requirements only and ignore the non-functional requirements for the time being. Thus, \textquoteleft\textit{immediate-func}\textquoteright refers to the satisfaction conditions for the functional requirements captured by that goal. Requirement analysts are required to provide the \textit{immediate-func} annotations for each goal. \textit{Cumulative-func} annotations of a goal represent the set of satisfaction conditions that are derived by accumulating the \textit{immediate-func} annotations of the goal tree that is rooted at that goal. 

The notion of accumulating lower level satisfaction conditions stems from the idea that sub-goals lower in a goal tree provide more detailed accounts of ways in which higher-level goals might be satisfied. Thus, there is value in propagating these satisfaction conditions up the goal tree to obtain annotations (attached to each goal) that provide more detailed (and complete) formal accounts of alternative ways in which a goal is actually being satisfied. A formal annotation of the goals empowers requirement analysts to use automated reasoners for consistency and compliance checking, thereby releasing the analysts from laborious and complex manual analysis and evaluations. The satisfaction reconciliation machinery can be viewed as a black box that takes context-independent goal satisfaction conditions as input and produces context-sensitive cumulative satisfaction conditions as output. The implications of the \textit{cumulative} satisfaction conditions, for a given i$^*$ model, depend on the precision with which analysts specify the \textit{immediate} satisfaction conditions \cite{PSEER,PCTk}.


Given a i$^*$ model configuration we may identify mainly two different types of conflicts that may exist within the model - \textit{entailment} and \textit{consistency}. The solution presented in this paper tries to answer the question - \textquotedblleft \textit{Given a i$^*$ modelconfiguration, how can we remove all conflicts existing within that i$^*$ model and generate a conflict-free configuration (version)?}\textquotedblright. Thus, given a i$^*$ model configuration, there exists a vast space of modified configurations that addresses this issue and generates conflict-free versions. It becomes quite infeasible for the analysts to enumerate the complete search space which, in turn, may result in analysts coming up with sub-optimal solutions to the i$^*$ model maintenance problem.


Goal models have AND-decompositions as well as OR-decompositions. AND-decompositions capture the lower level subgoals that must be fulfilled in order to satisfy a higher level goal. OR-decompositions, on the other hand, capture alternatives for fulfilling a given goal. Each OR-decomposition link shows one possible means for fulfilling the parent goal. This gives rise to the notion of goal subgraphs that define unique solutions for fulfilling high level goals. We call these subgraphs \textquotedblleft \textit{OR-refined goal models}\textquotedblright. Let us first define what we mean by OR-refined goal models.

\noindent\textbf{Definition.} \textit{OR-refined Goal Models} (ORGMod\textit{s}). An OR-refined goal model for a given high level goal $G$ is one with no OR-alternatives. It is obtained by committing to a specific OR-alternative wherever OR-alternatives exist in the goal model. 

$ORGMod$s can be derived by performing a modified depth-first-search (DFS) of the i$^*$ model subtree rooted at $G$ such that - 
\begin{list}{\textit{\roman{qcounter}}.~}{\usecounter{qcounter}}
	\item whenever we encounter an AND-decomposition, we include all the children in the decomposition, and
	\item whenever we encounter an OR-decomposition, we commit to only one of the possible alternatives.
\end{list}
\begin{figure}[!t]
	\centering
	\includegraphics[width=0.9\textwidth, page=1]{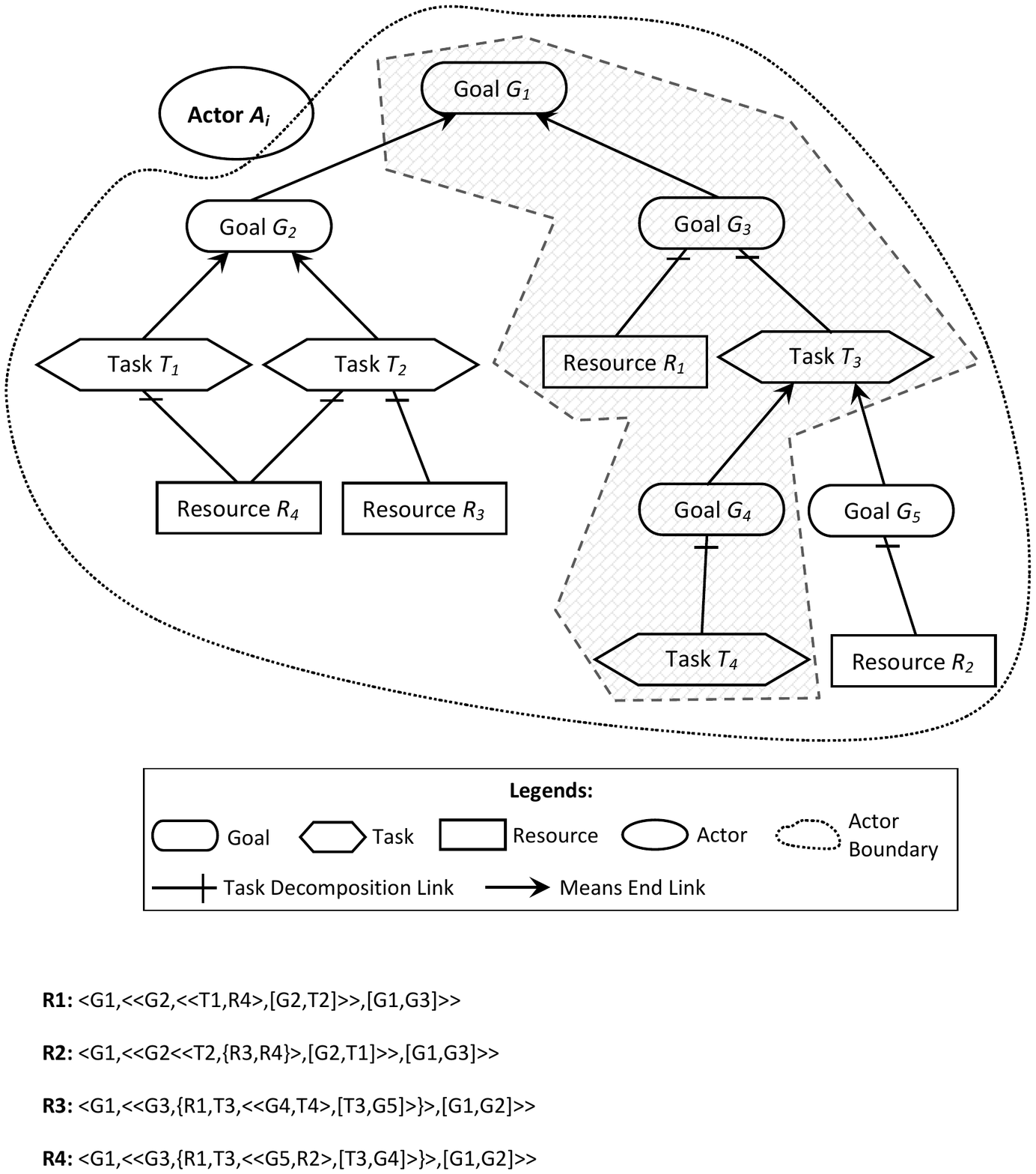}
	\caption{An OR-refined Goal Model highlighted within the i$^*$ model}
	\label{fig:str_rout}
\end{figure}

For instance, let us consider the high-level goal \textit{G$_1$} in actor \textit{A$_i$} (Figure \ref{fig:str_rout}). A possible $ORGMod$  for achieving this goal is marked by a dashed polygon. Assuming that actor \textit{A$_i$} does not depend on any other actor for successfully accomplishing task \textit{T$_4$}, the $ORGMod$  so identified can be written using a top-down approach as $\langle G_1,G_3,\{R_1,\langle T_3,G_4,T_4\rangle\}\rangle$. This notation is quite easy to follow. We start with the root artefact $G_1$ and keep tracing the i$^*$ model until we reach leaf-level artefacts. Sequences of artefacts within angle brackets \textquoteleft$\langle \rangle$\textquoteright  represent successive levels in the strategic routine whereas \textquoteleft\{\}\textquoteright are used to capture \textit{AND-decompositions}. So $\langle T_3,G_4,T_4\rangle$ represents three successive levels of the $ORGMod$ including task $T_3$ whereas $G_3,\{R_1,T_3\}$ represents an \textit{AND-decomposition} of goal $G_3$ into resource $R_1$ and task $T_3$. Both angle brackets and braces can be nested within one another.



The satisfaction reconciliation machinery processes the immediate satisfaction conditions of individual goals and builds their corresponding cumulative satisfaction conditions. We deploy this machinery between adjacent levels of an $ORGMod$. The cumulative satisfaction condition of task $T_4$ is combined with the immediate satisfaction condition of goal $G_4$ to obtain the cumulative satisfaction condition of goal $G_4$. The cumulative satisfaction condition of goal $G_4$ is then combined with the immediate satisfaction condition of task $T_3$ to obtain the cumulative satisfaction condition of task $T_3$ and so on until we reach the root goal $G_1$.

\subsection{$ORGMod$ Extraction}
As mentioned previously, it is cumbersome to derive the cumulative satisfaction condition for an entire i$^*$ model. Instead, the satisfaction reconciliation process can be restricted to one or more  $ORGMod$s. This is practically more useful as requirement analysts may wish to see the cumulative satisfaction conditions of some desired goals. $ORGMod$s represent goal sub-models that are derived from the original i$^*$ model and play a decisive role in the satisfaction reconciliation process.  $ORGMod$s help in pruning alternatives that can be excluded from the satisfaction reconciliation process.

\begin{figure}[!b]
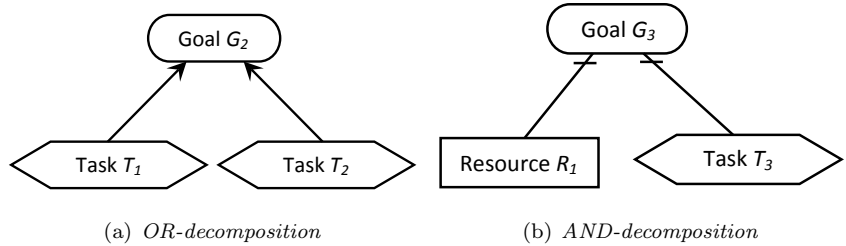

	\begin{center}
		\subfigure[\textit{OR-decomposition}]{\label{fig:SR-XOR}\includegraphics[width=0.45\textwidth, page=4]{figures.pdf}}\,\,
		\subfigure[\textit{AND-decomposition}]{\label{fig:SR-AND}\includegraphics[width=0.45\textwidth, page=5]{figures.pdf}} \\
	\end{center}
	\caption{Goal models illustrating the two different types of decompositions or splits that goals can undergo}
	\label{fig:SR-illustr}
\end{figure}

The pruning process also helps in assigning unique labels to $ORGMod$s. A routine label  identifies the order in which satisfaction conditions can be reconciled for a chosen high-level goal. Consider the \textit{OR-decomposition} shown in Figure \ref{fig:SR-XOR}. It captures two different alternates to achieve the high-level goal $G_2$. This results in two different labels for the two different $ORGMod$s that can be obtained:\\

\noindent Label 1:$\langle\langle G_2,T_1\rangle,[G_2,T_2]\rangle$\\
Label 2: $\langle\langle G_2,T_2\rangle,[G_2,T_1]\rangle$\\

Each routine label identifies an $ORGMod$ for satisfying goal $G_2$ and an \textit{exclusion set} (as defined in \cite{PSEER}). \textit{Exclusion sets} are used to list those alternate paths that were not selected at an \textit{OR-decomposition} for the given routine label. So for Label 1, goal $G_2$ is achieved successfully by performing task $T_1$ and, hence, [$G_2,T_2$] is in the exclusion set. $\langle G_2,T_1\rangle$ specifies  that satisfaction reconciliation must occur from $T_1$ to $G_2$. Similarly, we obtain Label 2 if we choose task $T_2$ over task $T_1$.

Consider the \textit{AND-decomposition} shown in Figure \ref{fig:SR-AND}. \textit{AND-decompositions} denote independent objectives that can be executed in any order or even in parallel. Sibling goals  are represented using \{\} in the $ORGMod$ sequence. Figure  \ref{fig:SR-AND} has the following label:\\

\noindent Label 3: $\langle\langle G_3,\{\langle R_1\rangle,\langle T_3\rangle\}\rangle,[\emptyset]\rangle$\\

$\{\langle R_1\rangle,\langle T_3\rangle\}$ represents independent resource and task contained within a set. This set is considered as a separate element in the outer sequence of goals. Also, the exclude set is null as \textit{AND-decompositions} do not provide choices to requirement analysts.

We intend to automate the process of extracting all possible $ORGMod$s that have a particular goal as root. This goal may be randomly chosen by the requirement analysts for cumulative satisfaction reconciliation. There are three sub-processes involved with the routine label extraction process - \textit{path traversal}, \textit{extracting decomposition sequences}, and \textit{deriving routine labels}. The following sections elaborate on these sub-processes.

\subsubsection{Path Traversal}
Any i$^*$ model can be viewed as a goal graph that already has an embedded tree structure. All the goals can be considered as generic nodes in a tree with all the decomposition links serving as edges. We ignore softgoals and contribution links for the time being. We perform a depth-first search on the goal graph. The path traversal process returns a list of paths, each of which is a sequence of goals from the root to the leaves and with no parallel edges. Applying the path traversal procedure on the i$^*$ model shown in Figure \ref{fig:str_rout} with goal $G_1$ as the chosen locus of satisfaction reconciliation, we get the following list of paths:\\

\noindent\textbf{Path List:}\\
\texttt{Path 1:}$\langle G_{1(X)},G_{2(X)},T_{1},R_{4}\rangle$\\
\texttt{Path 2:}$\langle G_{1(X)},G_{2(X)},T_{2(A)},R_{3}\rangle$\\
\texttt{Path 3:}$\langle G_{1(X)},G_{2(X)},T_{2(A)},R_{4}\rangle$\\
\texttt{Path 4:}$\langle G_{1(X)},G_{3(A)},R_{1}\rangle$\\
\texttt{Path 5:}$\langle G_{1(X)},G_{3(A)},T_{3(X)},G_{4},T_{4}\rangle$\\
\texttt{Path 6:}$\langle G_{1(X)},G_{3(A)},T_{3(X)},G_{5},R_{2}\rangle$\\

The symbols \textit{(X)} and \textit{(A)} are used to mark goals that undergo \textit{OR-decompositions} and \textit{AND-decompositions}, respectively. \textit{OR-decompositions} are analogous to exclusive gateways whereas \textit{AND-decompositions} are analogous to parallel gateways. These paths are collected and segmented into groups based on decomposition sequences.

\subsubsection{Extracting Decomposition Sequences}
\begin{figure}[h]
	\centering
	\includegraphics[width=0.7\textwidth, page=6]{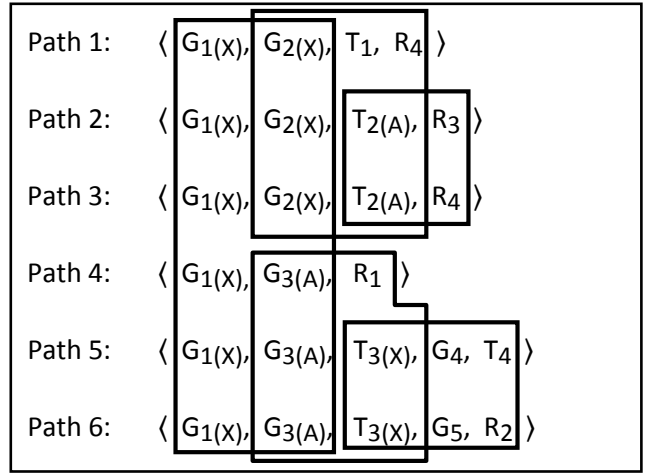}
	\caption{Decomposition sequence segmentation of the path list for the i$^*$ model in Figure \ref{fig:str_rout}.}
	\label{fig:dec_seq}
\end{figure}

\begin{figure}[!b]
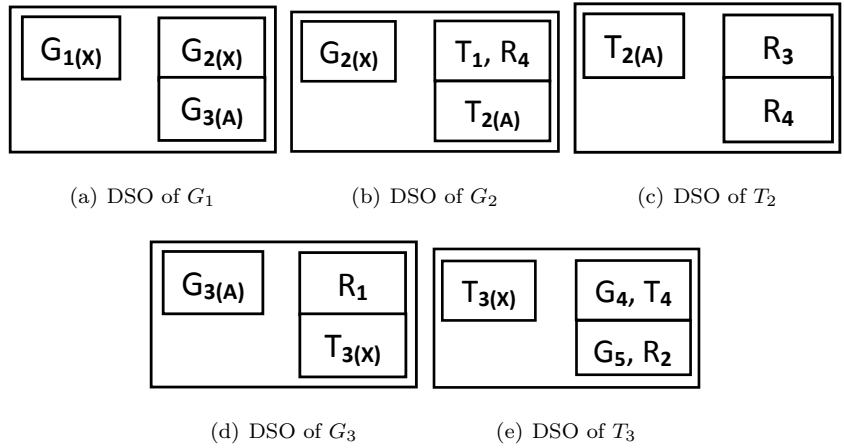

	\begin{center}
		\subfigure[DSO of $G_1$]{\label{fig:DSO-G1}\includegraphics[width=0.3\textwidth, page=7]{figures.pdf}}
		\subfigure[DSO of $G_2$]{\label{fig:DSO-G2}\includegraphics[width=0.3\textwidth, page=8]{figures.pdf}}
		\subfigure[DSO of $T_2$]{\label{fig:DSO-T2}\includegraphics[width=0.3\textwidth, page=9]{figures.pdf}}
		\subfigure[DSO of $G_3$]{\label{fig:DSO-G3}\includegraphics[width=0.3\textwidth, page=10]{figures.pdf}}
		\subfigure[DSO of $T_3$]{\label{fig:DSO-T3}\includegraphics[width=0.3\textwidth, page=11]{figures.pdf}}
	\end{center}
	\caption{Decomposition Sequence Objects (DSO) derived from the decomposition sequences in Figure \ref{fig:dec_seq}.}
	\label{fig:DSO}
\end{figure}

The path list derived in the previous section can be segmented into collections of goals as shown in Figure \ref{fig:dec_seq}. Each of these segments are referred to as decomposition sequences. These sequences can be captured using a generic format that represents objects of the decomposition class. Figure \ref{fig:DSO} shows the decomposition sequence objects (DSO(s)). Each decomposition sequence begins with a goal that undergoes an \textit{AND decomposition} or an \textit{OR decomposition} and its lower-level goals. Decomposition sequences either end at leaf-level goals or at the beginning of the next decomposition sequence.

$G_{1(X)}$ marks the beginning of the first decomposition sequence that ends in either $G_{2(X)}$ or $G_{3(A)}$. Both these decomposition artefacts mark the beginning of the next decomposition sequence. $G_{2(X)}$ can either end in the leaf-level artefact $R_{4}$ or in the decomposition artefact $T_{2(A)}$. Similarly, $G_{3(A)}$ either ends in the resource $R_{1}$ or the task decomposition node $T_{3(X)}$. This process is repeated to obtain a decomposition sequence for $T_{2(A)}$ and $T_{3(X)}$, respectively.

Each box in Figure \ref{fig:dec_seq} represents a decomposition sequence segment. Each segment represents a subsequence starting at some goal that either undergoes an \textit{OR-decomposition} or an \textit{AND-decomposition}. We can use these decomposition segments to derive Decomposition Sequence Objects (DSOs) as shown in Figure \ref{fig:DSO}. Each decomposition sequence begins with a different goal and can be mapped to a unique DSO for that goal. For example, Figure \ref{fig:DSO-G1} shows the DSO for goal $G_1$. For each DSO, the label on the left represents the root goal and the list of labels on the right represents the decomposition subsequences beginning at the root goal and ending at either a leaf-level goal or at another decomposition goal. The left-most decomposition sequence for goal $G_1$ (leftmost segment in Figure \ref{fig:dec_seq}) has two subsequences - one ending in $G_{2(X)}$ and the other ending in $G_{3(A)}$. The DSO for $G_{2(X)}$ represents two subsequences - one ending in $R_4$ and the other in $T_{2(A)}$.

Since goal models have an inherent tree structure, the parent-child relationship existing between the goals is preserved in the decomposition sequence objects as well. While developing $ORGMod$s, this relationship plays a vital role. For example,  sequences of the artefact $T_{3(X)}$ (in \ref{fig:DSO-T3}) must be resolved before creating the sequences of $G_{3(A)}$ (Figure \ref{fig:DSO-G3}). In general, decomposition sequences of the child goals must be incorporated within the parents. DSOs create a unique model that is independent of the original path list. This enables the creation of completely independent $ORGMod$s. Each DSO must have a method that accumulates the decomposition subsequences of the child goals and assembles them with the parent subsequence. 

\subsubsection{Deriving Routine Labels}
In order to derive all possible $ORGMod$s that stem from a particular goal (as desired by requirement analysts), we need to provide the first decomposition sequence only. Every decomposition sequence passes its own sequence to it's child decomposition sequence. The child decomposition sequence uses recursion to build all the possible subsequences and merges them with the parent sequence. The merging process is repeated backwards till we reach the first decomposition sequence. The final merge operation produces the set of all possible $ORGMod$ labels that can be derived for the given goal from the goal model. 

For the i$^*$ model of Figure \ref{fig:str_rout}, we begin with the decomposition sequence object of goal $G_1$. $G_1$ passes its subsequence to its child decomposition sequences $G_2$ and $G_3$. $G_2$ and $G_3$ build their own sequences separately and merge them with the parent sequence of $G_1$. The following sequence of operations illustrates the routine label derivation process. The symbol $\rightarrow$ is used to indicate the passing of sequences from a parent sequence to a child sequence. The symbol $\leftarrow$ is used to denote the passing of all merged sequences from the child to the parent. The beginning subsequence before the first decomposition in our example is $\emptyset$ which calls the first decomposition sequence.\\

\noindent\framebox{$\emptyset\rightarrow G_{1(X)}$}\\

In order to generate the decomposition sequences, we look at the DSO of $G_{1(X)}$ (Figure \ref{fig:DSO-G1}). $G_{1(X)}$ must process its child decomposition sequences before returning the final list of routine labels.\\

\noindent\framebox{$\langle G_{1(X)},\langle G_{2(X)}\rangle\rangle$}\\
\noindent\framebox{$\langle G_{1(X)},\langle G_{3(A)}\rangle\rangle$}\\

$G_1$ passes its subsequence ($\emptyset$) to $G_{2(X)}$ and $G_{3(A)}$.\\

\noindent\framebox{$\emptyset\rightarrow G_{2(X)}$}\\
\noindent\framebox{$\emptyset\rightarrow G_{3(A)}$}\\

$G_{2(X)}$ processes its child subsequences before returning the final list of sequences to $G_{1(X)}$. Child subsequences are obtained from the DSO of $G_{2(X)}$, shown in Figure \ref{fig:DSO-G2}.\\

\noindent\framebox{$\langle G_{2(X)},\langle T_1,R_4\rangle\rangle$}\\
\noindent\framebox{$\langle G_{2(X)},\langle T_{2(A)}\rangle\rangle$}\\

$G_{2(X)}$ passes its its child subsequence ($\emptyset$) to $T_{2(A)}$ which subsequently processes its child sequences as listed in it's DSO (Figure \ref{fig:DSO-T2}).\\

\noindent\framebox{$\emptyset\rightarrow T_{2(A)}$}\\
\noindent\framebox{$\langle T_{2(A)},\{\langle R_3\rangle,\langle R_4\rangle\}\rangle$}\\

All the child subsequences of $G_{2(X)}$ are sent back to $G_{2(X)}$ and combined with it's own sebsequence.\\

\noindent\framebox{$G_{2(X)}\leftarrow\langle T_1,R_4\rangle$}\\
\noindent\framebox{$G_{2(X)}\leftarrow\langle T_{2(A)},\{\langle R_3\rangle,\langle R_4\rangle\}\rangle$}\\

Similarly, $G_{3(A)}$ also processes its child subsequences before returning the final list of sequences to $G_{1(X)}$. Figure \ref{fig:DSO-G3} shows the list child subsequences that must be processed:\\

\noindent\framebox{$\langle G_{3(A)},\{\langle R_1\rangle,\langle T_{3(X)}\rangle\}\rangle$}\\

We proceed in a similar manner and process the child sequences of $T_{3(X)}$ (shown in Figure \ref{fig:DSO-T3}). These subsequences are returned to their parent $G_{3(A)}$ as follows:\\

\noindent\framebox{$G_{3(A)}\leftarrow\{\langle R_1\rangle,\langle T_{3(X)},\langle G_4,T_4\rangle\rangle\}$}\\
\noindent\framebox{$G_{3(A)}\leftarrow\{\langle R_1\rangle,\langle T_{3(X)},\langle G_5,R_2\rangle\rangle\}$}\\

Both $G_{2(X)}$ and $G_{3(A)}$ merge their child sequences with their own subsequence and return the resulting sequences to their parent $G_{1(X)}$.\\

\noindent\framebox{$G_{1(X)}\leftarrow\langle G_{2(X)},\langle T_1,R_4\rangle\rangle$}\\
\noindent\framebox{$G_{1(X)}\leftarrow\langle G_{2(X)},\langle T_{2(A)},\{\langle R_3\rangle,\langle R_4\rangle\}\rangle\rangle$}\\
\noindent\framebox{$G_{1(X)}\leftarrow\langle G_{3(A)},\{\langle R_1\rangle,\langle T_{3(X)},\langle G_4,T_4\rangle\rangle\}\rangle$}\\
\noindent\framebox{$G_{1(X)}\leftarrow\langle G_{3(A)},\{\langle R_1\rangle,\langle T_{3(X)},\langle G_5,R_2\rangle\rangle\}\rangle$}\\

$G_{1(X)}$ receives these sequences from its child decomposition sequences and combines them with it's own subsequence (in this case $\emptyset$) and generates the set of all strategic routine labels that can be derived for goal $G_{1(X)}$.\\

\noindent\framebox{$\langle G_{1(X)},\langle G_{2(X)},\langle T_1,R_4\rangle\rangle\rangle$}\\
\noindent\framebox{$\langle G_{1(X)},\langle G_{2(X)},\langle T_{2(A)},\{\langle R_3\rangle,\langle R_4\rangle\}\rangle\rangle\rangle$}\\
\noindent\framebox{$\langle G_{1(X)},\langle G_{3(A)},\{\langle R_1\rangle,\langle T_{3(X)},\langle G_4,T_4\rangle\rangle\}\rangle\rangle$}\\
\noindent\framebox{$\langle G_{1(X)},\langle G_{3(A)},\{\langle R_1\rangle,\langle T_{3(X)},\langle G_5,R_2\rangle\rangle\}\rangle\rangle$}\\

\subsubsection*{Result}
When the process concludes, we obtain the list of all possible $ORGMod$ labels from the decomposition sequence of $G_{1(X)}$. In the previous illustrations, we have omitted \textit{exclusion sets} for the sake of simplicity. Exclusion sets are obtained when child sequences are assembled by the parent sequence. The final list of $ORGMod$ labels, including exclusion sets, are as follows:\\

\noindent\texttt{Routine 1:}\framebox{$\langle G_{1},\langle\langle G_{2},\langle\langle T_1,R_4\rangle,[G_2,T_2]\rangle\rangle,[G_1,G_3]\rangle\rangle$}\\
\noindent\texttt{Routine 2:}\framebox{$\langle G_{1},\langle\langle G_{2},\langle\langle T_{2},\{\langle R_3\rangle,\langle R_4\rangle\}\rangle,[G_2,T_1]\rangle\rangle,[G_1,G_3]\rangle\rangle$}\\
\noindent\texttt{Routine 3:}\framebox{$\langle G_{1},\langle\langle G_{3},\{\langle R_1\rangle,\langle T_{3},\langle\langle G_4,T_4\rangle,[T_3,G_5]\rangle\rangle\}\rangle,[G_1,G_2]\rangle\rangle$}\\
\noindent\texttt{Routine 4:}\framebox{$\langle G_{1},\langle\langle G_{3},\{\langle R_1\rangle,\langle T_{3},\langle\langle G_5,R_2\rangle,[T_3,G_4]\rangle\rangle\}\rangle,[G_1,G_2]\rangle\rangle$}\\

\begin{figure}[h]
	\centering
	\includegraphics[width=0.9\textwidth, page=12]{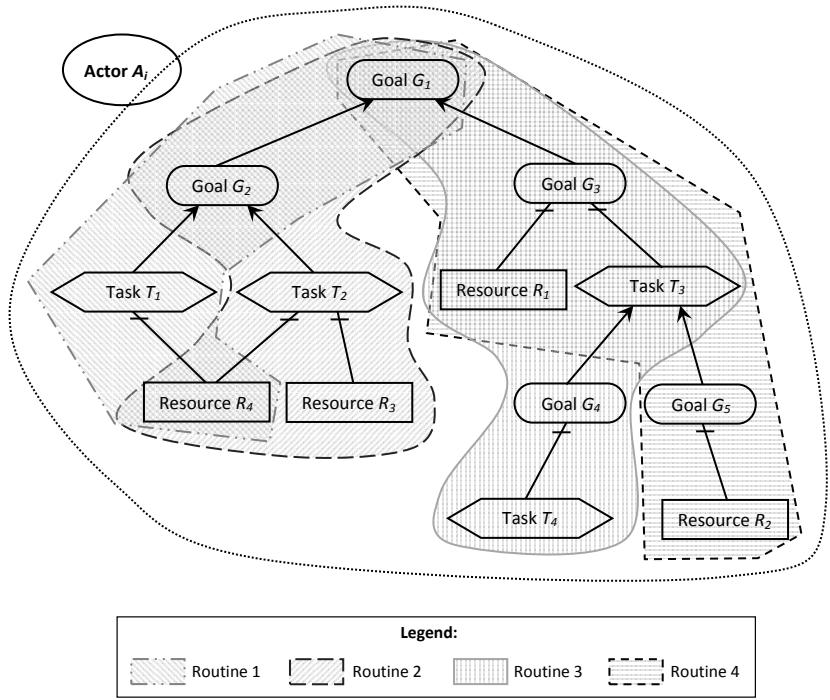}
	\caption{The four possible \textit{$ORGMod$s} that have been derived for goal $G_1$, highlighted within the i$^*$ model of actor \textit{A$_i$}}
	\label{fig:str_rout_all}
\end{figure}

Figure \ref{fig:str_rout_all} highlights the four different $ORGMod$s that have been derived using the process described above. Consider the first $ORGMod$. At $G_1$ we choose the path $G_1,G_2$ and, hence, [$G_1,G_3$] becomes the exclusion set. Similarly, at $G_2$, we choose $T_1$ over $T_2$ and, hence, [$G_2,T_2$] becomes the exclusion set for this point.

The process described above is exhaustive and generates all possible $ORGMod$ labels that can be derived for any desired goal. Each such routine label consists of a list of goals that can exist in sequences, exclusion sets (\textit{OR-decompositions}) or parallel sets (\textit{AND-decompositions}). Once we obtain the routine labels corresponding to some goal, we can extract the \textit{immediate-func} annotations and reconcile them to derive the \textit{cumulative-func} annotations for individual goals. The process of satisfaction reconciliation has been elaborated in the next section. 

\subsection{Satisfaction Reconciliation Operators}
The \textit{satisfaction reconciliation} operation takes the goal satisfaction conditions of the higher level goal and cumulative satisfaction conditions of the lower level goals as input and determines the cumulative satisfaction conditions of the higher level goal as output. This definition implies that, for leaf-level goals, the goal and cumulative satisfaction condition are identical, provided they are not dependant on leaf level goals residing in other actor boundaries. Consider the following simple example, where we have the higher level goal \textit{Make\_Payment} which decomposes to the lower level goals \textit{Check\_Balance} and \textit{Transfer\_Funds}. Assuming informal annotations of satisfaction conditions, we may have a satisfaction annotation scenario as follows:
\begin{enumerate}
	\item \textit{Check\_Balance}: \texttt{Insufficient}
	\item \textit{Transfer\_Funds}: \texttt{Cheque\_bounced}
	\item \textit{Make\_Payment}: \texttt{Payment\_done}
\end{enumerate}
Clearly this results in an inconsistent satisfaction scenario. Background rules may be defined on the underlying knowledge base to prevent such inconsistent satisfaction conditions from occurring in the same satisfaction scenario. We assume that every i$^*$ model has an underlying knowledge base (KB) that provides the basis for preventing such inconsistencies.

Assuming that goal satisfaction conditions are represented as sets of canonical, non-redundant clauses in conjunctive normal form (CNF) \cite{21-PSEER}\cite{PCTk}\cite{PSEER}, we represent the satisfaction condition of a goal $A$ as the set \textit{IE}($A$)=\{$ie_1,ie_2,...,ie_n$\} and its derived cumulative satisfaction condition as the set \textit{CE}($A$)=\{$ce_1,ce_2,...,ce_m$\}, for a given $ORGMod$. Let $A_i$ and $A_j$ be two adjacent level goals in the i$^*$ model. We define \textit{rec}($A_i$,$A_j$) as the satisfaction reconciliation operation that derives the cumulative satisfaction condition of $A_i$ by combining it's immediate satisfaction conditions with the cumulative satisfaction conditions of $A_j$, i.e., \textit{CE}($A_i$)=\textit{rec}($A_i$,$A_j$). While deriving the cumulative satisfaction conditions of any artefact, we consider two components:
\begin{itemize}
	\item all immediate satisfaction conditions of that goal that are derivable from the lower level goal(s), and
	\item all additional satisfaction conditions of the lower level goal(s) whose negations are not listed in the immediate satisfaction conditions of that goal.
\end{itemize} 

Thus, we define the satisfaction reconciliation operation as follows:
\begin{equation}
\label{rec}
CE(A_i)= rec(A_i,A_j)\nonumber
\end{equation}
\begin{equation}
=\{IE(A_i)\cap CE(A_j)\} \bigcup\{CE(A_j)\setminus \neg IE(A_i)\},\\
\end{equation}
such that $\neg IE(A_i)=\{\neg ie_1,\neg ie_2,...,\neg ie_n\}$ and \textit{CE}($A_i$) contains all immediate satisfaction conditions of $A_i$ that are derivable from $A_j$ (given by the intersection operation), as well as all additional satisfaction conditions of $A_j$ whose negation are not contained in \textit{IE}($A_i$) (given by the set difference operation). 

If consistency is not satisfied for all members of $CE(A_j)$, then we proceed to include as many cumulative satisfaction conditions of $A_j$ as possible while maintaining consistency with the knowledge base KB. The following example provides a better understanding of the reconciliation operation. 

\textbf{Example:} Let $A$ and $B$ be adjacent level goals in an i$^*$ model where $B$ is the next lower level below $A$. Let $IE(A)$=\{$a,b,c$\} and $CE(B)$=\{$\neg b,c,d,e$\}. The cumulative satisfaction condition of $A$, denoted by $CE(A)$ can be evaluated as follows:

\begin{align*}
IE(A) &= \{a,b,c\}\\
CE(B) &= \{\neg b,c,d,e\}\\
CE(A) &= rec(A,B)\\
&= \{IE(A)\cap CE(B)\}\bigcup\\&\{CE(B)\setminus \neg IE(A)\} \\
IE(A)\cap CE(B) &= \{c\},\\
\neg IE(A) &= \{\neg a,\neg b,\neg c\},\\
CE(B)\setminus \neg IE(A) &= \{\neg b,c,d,e\}\setminus\{\neg a,\neg b,\neg c\}\\
&= \{c,d,e\}
\\
\therefore CE(A) &= \{c\} \cup \{c,d,e\}\} = \{c,d,e\}.
\end{align*}

For finding out the cumulative satisfaction condition of any two model artefacts $A$ and $B$ (given by Eqn. \ref{rec}), the intersection operation is redundant, i.e., the satisfaction reconciliation operation $rec(A,B)$ can be represented as - 
\begin{equation}
\label{eq:rec}
	CE(A)=rec(A,B)=\{CE(B)\setminus \neg IE(A)\}
\end{equation}
We prove equation \ref{eq:rec} by establishing that $IE(A)\cap CE(B) \subseteq CE(B)\setminus \neg IE(A)$, and, hence, redundant. For any two satisfaction condition sets $X$ and $Y$ the universe of satisfaction conditions for their satisfaction reconciliation operation is given by $U=X\cup Y\cup\neg X\cup\neg Y$ where $\neg X$ and $\neg Y$ can be derived by negating each satisfaction condition within the set, as shown in Eqn. \ref{rec}. Also, we know that \textit{set difference} between two sets $X$ and $Y$ can be expressed using \textit{set intersection} as $X\setminus Y = X\cap Y^C$ where $Y^C$ represents the compliment set of $Y$ with respect to the universe $U$.
\begin{align*}
\because CE(B)\setminus \neg IE(A) &= CE(B)\cap(\neg IE(A))^C,\\
\text{R.T.P.:- }CE(B)\cap IE(A) &\subseteq CE(B)\cap(\neg IE(A))^C\\
\Rightarrow \text{R.T.P.:-}IE(A) &\subseteq (\neg IE(A))^C.
\end{align*}

For any two satisfaction condition sets $X$ and $Y$ along with their universe $U$, the set $(\neg X)^C$ consists of the union of three different subsets - 
\begin{enumerate}
	\item All elements of the set $X$.
	\item All elements in $Y$ that are not in $\neg X$, i.e., $Y\setminus \neg X$.
	\item All elements in $\neg Y$ that are not in $\neg X$, i.e., $\neg Y\setminus\neg X$.
\end{enumerate}

Thus, we have the following derivation:
\begin{align*}
(\neg X)^C &= X \cup (Y\setminus\neg X) \cup (\neg Y\setminus\neg X)\\
&= X \cup (Y\cap\neg X^C) \cup (\neg Y\cap\neg X^C)\\
&= X \cup \{(Y\cup\neg Y)\cap\neg X^C\}\hspace{1cm} \text{[Distributive Law]}
\end{align*}

If the set $\{(Y\cup\neg Y)\cap\neg X^C\} = \emptyset$, then $(\neg X)^C = X$; else $X\subset(\neg X)^C$. Thus, we can conclude that $X\subseteq(\neg X)^C$. Replacing the set $X$ with $IE(A)$ and the set $Y$ with $CE(B)$, we can conclude that - 
\begin{align*}
IE(A) &\subseteq (\neg IE(A))^C\\
\Rightarrow CE(B)\cap IE(A) &\subseteq CE(B)\cap(\neg IE(A))^C\\
\Rightarrow CE(B)\cap IE(A) &\subseteq \ CE(B)\setminus \neg IE(A).\\
\therefore \text{from Eqn. 1, } CE(A)=rec(A,B)&=\{CE(B)\setminus \neg IE(A)\}.
\end{align*}


Goal decompositions have multiple children in most cases, and this requires the satisfaction reconciliation machinery to be more generic. We need to explicitly define the mechanisms when there are multiple subgoals resulting from a given goal decomposition. This becomes mandatory for goals that undergo \textit{AND-decomposition}. However, for a goal undergoing \textit{OR-decomposition}, the interpretation changes. As already stated, we perform satisfaction reconciliation on a \textit{per $ORGMod$} basis. This implies that we select only one alternative whenever we encounter an OR-decomposition. However, requirement analysts may desire to answer questions like -  \textit{\textquotedblleft Do I have a strategy that works properly?\textquotedblright} In order to answer such questions, and for completeness of the framework, we need to specify the machinery for satisfaction reconciliation over OR-decompositions as well.

\subsubsection{AND-reconciliation Operator}
\label{ANDmerge}
\textit{AND-decompositions} can occur within an $ORGMod$ itself. The example shown in Figure \ref{fig:str_rout} illustrates an \textit{AND-decomposition} of goal $G_3$ into resource $R_1$ and task $T_3$. Using our bottom-up approach, the cumulative satisfaction conditions of $R_1$ and $T_3$ will be evaluated first. Since $R_1$ is a leaf-level resource, \textit{CE}($R_1$)=\textit{IE}($R_1$). The cumulative satisfaction condition of goal $T_3$ is obtained from the next level goal $G_4$ as \textit{CE}($T_3$)=$rec(T_3,G_4)$. The \textit{AND-reconciliation} operation will be required when we try to evaluate the cumulative satisfaction conditions of the goal $G_3$ using $CE(R_1)$ and $CE(T_3)$.

Let $A_j$ and $A_k$ be two goals that result from the \textit{AND-decomposition} of the higher-level goal $A_i$. Let \textit{CE}($A_j$)=\{$ce_{j1},ce_{j2},...,ce_{jm}$\} and \textit{CE}($A_k$)=\{$ce_{k1},ce_{k2},...,ce_{kn}$\}, respectively, where $ce_{xy}$ represents the $y$-th member in the cumulative satisfaction condition of goal $x$. Let the goal satisfaction conditions of $A_i$ be denoted as $IE(A_i)$. In that case, we define \textit{CE}(A$_i$) using the \textit{ANDrec}() satisfaction reconciliation operation as follows: 
\begin{equation}
\label{ANDrec_bas}
\textit{ANDrec}(A_i,A_j,A_k)=\{\textit{rec}(A_i,A_j)\bigcup \textit{rec}(A_i,A_k)\}.
\end{equation}

In general, if a goal $A_N$ undergoes an \textit{AND-decomposition} to generate the set of goals $A_1,A_2,...,A_K$, then we can define \textit{CE}($A_N$) using the \textit{ANDrec}() operation as follows:
\begin{equation}
\label{eq:ANDrec}
\textit{ANDrec}(A_N,A_1,A_2,...,A_k)=\left\lbrace \bigcup_{P=1}^K\textit{rec}(A_N,A_P)\right\rbrace.
\end{equation}

\noindent\textbf{Guard condition.} Since goal models are sequence agnostic, the ordering of sibling events that stem from an AND-decomposition is abstracted from the model description. Thus, a correct i$^*$ model design demands that, during runtime, the order of executing events should not impact the state of the system. For example, with respect to eqn. \ref{ANDrec_bas}, the system should reach the same state of affairs if $A_j$ is performed before $A_k$ or \textit{vice-versa}. This property is known as \texttt{"Commutativity of State Updation"}. Let \textit{State\_Updt}() denote the state updation operator such that \textit{State\_Updt}($A_p$) results in changing the current state of the system by incorporating the satisfaction conditions of performing event $A_p$ and obtaining a new state. Thus, with respect to eqn. \ref{ANDrec_bas}, the commutativity property demands that-
\begin{equation}
\label{sttupd_bas}
\textit{State\_Updt}(A_j,\textit{State\_Updt}(A_k))=
\textit{State\_Updt}(A_k,\textit{State\_Updt}(A_j))
\end{equation}
In general, if a goal $A_N$ undergoes an \textit{AND-decomposition} to generate the set of goals $A_1,A_2,...,A_K$, then commutativity is satisfied if applying the state updation operator on any random ordering of these $K$ events, results in the same final state. Commutativity of State Updation must be satisfied for AND-decompositions. For OR-decompositions, we need not worry about commutativity as they represent alternate strategies and we perform analysis on a per $ORGMod$ basis.

\subsubsection{OR-reconciliation Operator}
\label{ORmerge}
\begin{figure}[t]
	\centering
	\includegraphics[width=0.9\textwidth, page=2]{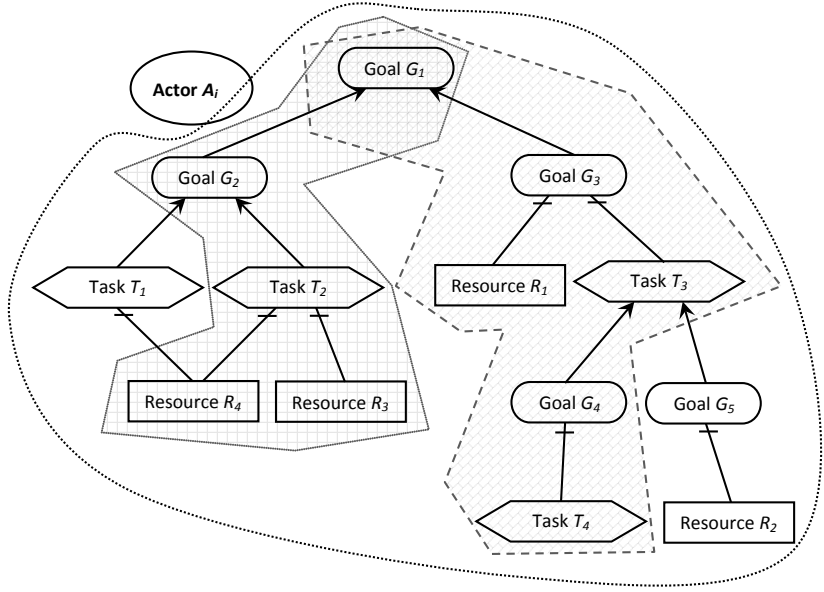}
	\caption{Two \textit{$ORGMod$s} highlighted for goal $G_1$ which undergoes a \textit{OR-decomposition} within the i$^*$ model of actor \textit{A$_i$}}
	\label{fig:2_str_rout}
\end{figure}

An \textit{OR-decomposition} provides alternate strategies for achieving the same goal. Since a $ORGMod$ chooses one particular alternative whenever it encounters an \textit{OR-decomposition}, we need an \textit{OR-reconciliation} whenever we want to combine the cumulative satisfaction conditions of two or more $ORGMod$s (or subroutines) at the point of an \textit{OR-decomposition}. Figure \ref{fig:2_str_rout} illustrates an example where we have two alternate $ORGMod$s for satisfying the goal $G_1$. The two $ORGMod$s are denoted as $\langle G_{1},\langle G_{3},\{\langle R_1\rangle,\langle T_3,\langle G_4,T_4\rangle\rangle\}\rangle\rangle$ and $\langle G_{1},\langle G_{2},\langle T_2,\{\langle R_3\rangle,\langle R_4\rangle\}\rangle\rangle\rangle$.

Let $A_j$ and $A_k$ be two goals that result from the \textit{OR-decomposition} of the higher-level goal $A_i$ and represent two different strategies for satisfying $A_i$. Let \textit{CE}($A_j$)=\{$ce_{j1},ce_{j2},...,ce_{jm}$\} and \textit{CE}($A_k$)=\{$ce_{k1},ce_{k2},...,ce_{kn}$\}, respectively. Let the goal satisfaction conditions of $A_i$ be denoted as $IE(A_i)$. In that case, we define \textit{CE}(A$_i$) using the \textit{ORrec}() satisfaction reconciliation operation as follows: 
\begin{equation}
\textit{ORrec}(A_i,A_j,A_k)=\{\{rec(A_i,A_j)\},\{rec(A_i,A_k)\}.
\end{equation}

In general, if a goal $A_N$ undergoes an \textit{OR-decomposition} to generate the set of goals $A_1,A_2,...,A_K$, then we can define \textit{CE}($A_N$) using the \textit{ORrec}() operation as follows:

\setlength{\arraycolsep}{0.0em}
\begin{eqnarray}
\label{eq:ORrec}
\textit{ORrec}(A_N,A_1,\cdots,A_K)&=&\lbrace rec(A_N,A_x) \rvert\nonumber\\ 
&&\forall x, A_x\in\{A_1,\cdots,A_K\}\rbrace.
\end{eqnarray}
\setlength{\arraycolsep}{5pt}

\subsubsection{Dependency Reconciliation Operator}
The previous examples, shown in Figures \ref{fig:str_rout} and \ref{fig:2_str_rout}, have an underlying assumption that all leaf-level artefacts are primitively satisfiable by the actor $A_i$. This means that the leaf-level artefacts do not depend on other actors for satisfying them and, hence, the $ORGMod$ shown in these examples are restricted to within the actor boundary. However, this may not be the case always.

\begin{figure}[t]
	\centering
	\includegraphics[width=\textwidth, page=3]{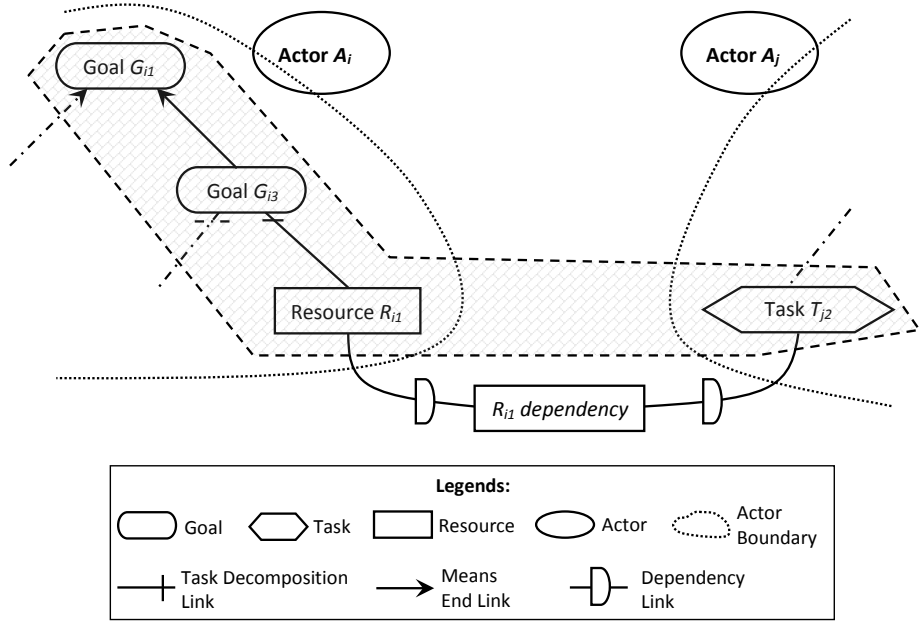}
	\caption{A \textit{strategic routine} that extends beyond the boundary of actor $A_i$ as the resource $R_{i1}$ depends on task $T_{j2}$ of actor $A_j$.}
	\label{fig:str_rout_dep}
\end{figure}

Consider the multi-actor SR-model shown in Figure \ref{fig:str_rout_dep}. The figure illustrates a strategic routine that spans beyond the actor boundary of $A_i$ and forays into the boundary of actor $A_j$. The \textit{$R_{i1}$ dependency} captures the requirement that actor $A_i$ needs to acquire the resource $R_{i1}$ and depends on actor $A_j$ to perform task $T_{j2}$ to provide the required resource. Unlike the previous examples, \textit{CE}($R_{i1}$)$\neq$\textit{IE}($R_{i1}$). In general, we can refine the formula for evaluating the cumulative effects of leaf level artefact $L_a$ as follows:

\begin{equation}
\label{DEPrec}
CE(L_a) = DEPrec(L_a) = \begin{cases}
IE(L_a), \hfill\text{if L$_a$ is independent}.\\
rec(L_a,L_b), \hspace{0.4cm}\text{using eqn.(\ref{rec}), if $L_a$ depends on $L_b$}.
\end{cases}
\end{equation}

Assuming an i$^*$ standard that, for any dependency, both the depender and the dependee are leaf-level artefacts and do not undergo any further decompositions, we proceed to derive the cumulative effect \textit{CE}($R_{i1}$) using the dependency effect reconciliation operation \textit{DEPrec}() as follows:
\begin{equation}
CE(R_{i1})=\textit{DEPrec}(R_{i1})=rec(R_{i1},T_{j2}). \nonumber
\end{equation}
such that $CE(T_{j2})\cap \neg IE(R_{i1})=\emptyset$, i.e., $IE(R_{i1})$ and $CE(T_{j2})$ are mutually consistent. If the intersection results in a non-empty set, then the corresponding dependency gives rise to inconsistencies within the model.

In general, we may have a chain of dependencies (not a cycle or loop) for satisfying a leaf-level i$^*$ artefact. Assuming that there are no cycles, let $\langle A_1,M_1\rangle\rightarrow\langle A_2,M_2\rangle\rightarrow\cdots\rightarrow\langle A_k,M_k\rangle$ represent a chain of transitive dependencies where $\langle A_i,M_i\rangle\rightarrow\langle A_j,M_j\rangle$ implies that a leaf-level artefact $M_i$ in actor $A_i$ is dependent on another leaf-level artefact $M_j$ in actor $A_j$. The cumulative effect annotation for the first artefact in the dependency chain, \textit{CE}($M_1$), can be derived from the following recurrence relation, using equation \ref{DEPrec}:
\begin{equation}
\label{eq:DEPrec_mult}
CE(M_i) = DEPrec(M_i) = \begin{cases}
IE(M_i), \hfill\text{if i=k}.\\
rec(M_i, DEPrec(M_{i+1})), \hspace{0.7cm}\text{$\forall i, 1\leq i\leq(k-1)$}.
\end{cases}
\end{equation}

\subsection{Illustrative Examples}

Let us illustrate the working of the  satisfaction reconciliation formalism with the help of some illustrative examples. We consider three different types of functional checks on an annotated i$^*$ model - namely \textit{entailment, consistency,} and \textit{minimality}. Of these, satisfying \textit{entailment} and \textit{consistency} is mandatory as they ensure the correctness of an i$^*$ model. \textit{Minimality} is an optional check that does not result in incorrect system states. We illustrate four different example that demonstrate different degrees of correctness for i$^*$ models.

\paragraph*{Case 1: Entailment not satisfied but Consistency satisfied}

\begin{figure}[h]
	\centering
	\includegraphics[width=0.9\textwidth, page=13]{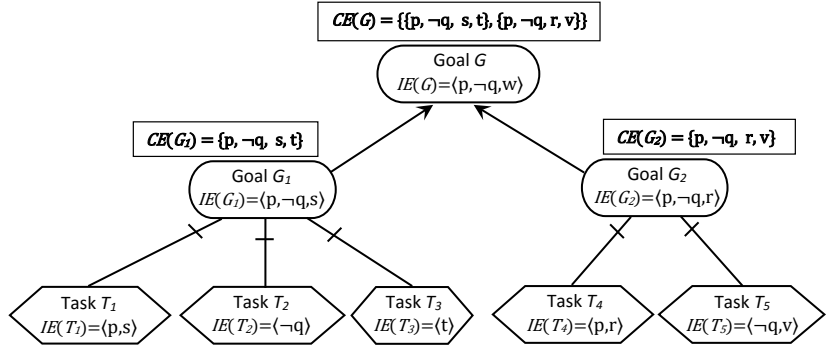}
	\caption{An illustrative example showing how  satisfaction reconciliation can be used to detect problems in \textit{entailment} although \textit{consistency} is ensured.}
	\label{fig:ec01}
\end{figure}

Consider the i$^*$ model shown in Figure \ref{fig:ec01}. It consists of a primary goal \textit{G} that undergoes an \textit{OR}-decomposition into goals \textit{G$_1$} and \textit{G$_2$}, which further undergo \textit{AND}-decompositions.

Every goal has been labelled with their goal satisfaction conditions, as specified by the requirement analysts. We now perform a  satisfaction reconciliation over this i$^*$ model using equations \ref{eq:ANDrec} and \ref{eq:ORrec}, defined in previous sections. The cumulative satisfaction conditions of goal \textit{G$_1$} is obtained using equation \ref{eq:ANDrec} as - 
\begin{align*}
CE(G_1) &= ANDrec(G_1,T_1,T_2,T_3)\\
&= \{rec(G_1,T_1) \cup rec(G_1,T_2) \cup rec(G_1,T_3)\} \\
rec(G_1,T_1) &= \{\{p, s\} \cup \emptyset\} = \{p,s\}, \hspace{1cm} \text{and } [CE(T_1)\cap \neg IE(G_1)=\emptyset]\\
rec(G_1,T_2) &= \{\{\neg q\} \cup \emptyset\} = \{\neg q\}, \hspace{1.2cm}\text{and }[CE(T_2)\cap \neg IE(G_1)=\emptyset]\\
rec(G_1,T_3) &= \{\emptyset \cup \{t\} = \{t\}, \hspace{1.94cm}\text{and }[CE(T_3)\cap \neg IE(G_1)=\emptyset]\\
\\
\therefore CE(G_1) &= \{\{p,s\} \cup \{\neg q\} \cup \{t\}\} = \{p,\neg q,s,t\}
\end{align*}
This has been shown in the figure as a label outside goal \textit{G$_1$}. Since $IE(G_1) \subseteq CE(G_1)$, this \textit{AND}-decomposition satisfies both \textit{entailment} and \textit{consistency}. Similarly, we proceed to evaluate the cumulative satisfaction conditions of goal \textit{G$_2$} as - 
\begin{align*}
CE(G_2) &= ANDrec(G_2,T_4,T_5)\\
&= \{p,\neg q,r,v\}
\end{align*}
This cumulative satisfaction condition has also been shown in the figure outside goal \textit{G$_2$}. Again, since $IE(G_2) \subseteq CE(G_2)$, this \textit{AND}-decomposition also satisfies both \textit{entailment} and \textit{consistency}. Following the bottom-up approach, we now proceed to evaluate the cumulative satisfaction conditions of goal \textit{G} using equation \ref{eq:ORrec} as follows - 
\begin{align*}
CE(G) &= ORrec(G,G_1,G_2)\\
&= \{\{p,\neg q,s,t\},\{p,\neg q,r,v\}\}
\end{align*}
The cumulative satisfaction conditions of goal \textit{G} has been labelled in the figure. Since neither of the members in $CE(G_1)$ or $CE(G_2)$  have any mutually conflicting satisfaction conditions with $IE(G)$, hence, we conclude that the satisfaction reconciliation is \textit{consistent}. However, since both $IE(G) \nsubseteq CE(G_1)$ and $IE(G) \nsubseteq CE(G_2)$, we conclude that none of the strategies ensure \textit{entailment}.

\paragraph*{Case 2: Entailment satisfied but Consistency is not}

\begin{figure}[h]
	\centering
	\includegraphics[width=0.9\textwidth, page=15]{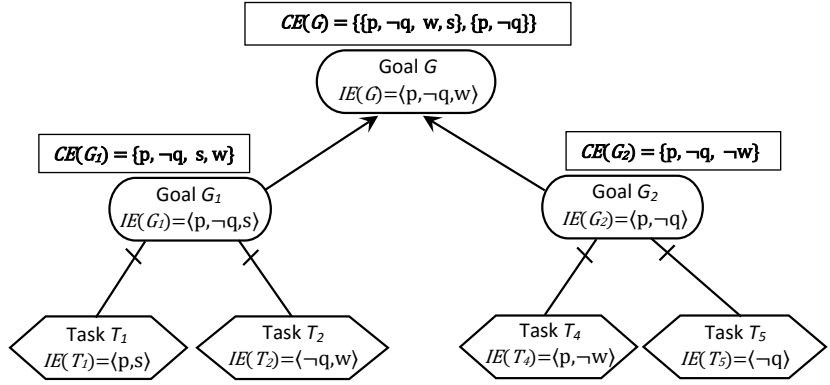}
	\caption{An example showing a \textit{consistency} conflict between the immediate annotations of parent goal $G$ and the cumulative annotations of child goal $G_2$.}
	\label{fig:ec10}
\end{figure}

Consider the i$^*$ model shown in Figure \ref{fig:ec10}. It consists of a primary goal \textit{G} that undergoes an \textit{OR}-decomposition into goals \textit{G$_1$} and \textit{G$_2$}, which further undergo \textit{AND}-decompositions.

Every goal has been labelled with their goal satisfaction conditions, as specified by the requirement analysts. We now perform a satisfaction reconciliation over this i$^*$ model using equations \ref{eq:ANDrec} and \ref{eq:ORrec}, defined in previous sections. The cumulative satisfaction conditions of goal \textit{G$_1$} is obtained using equation \ref{eq:ANDrec} as - 
\begin{align*}
CE(G_1) &= ANDrec(G_1,T_1,T_2)\\
&=\{p,\neg q,s,w\}
\end{align*}
Since $IE(G_1) \subseteq CE(G_1)$, this \textit{AND}-decomposition satisfies both \textit{entailment} and \textit{consistency}. Similarly, we proceed to evaluate the cumulative satisfaction condition of goal \textit{G$_2$} as - 
\begin{align*}
CE(G_2) &= ANDrec(G_2,T_4,T_5)\\
&= \{p,\neg q,\neg w\}
\end{align*}
Again, since $IE(G_2) \subseteq CE(G_2)$, this \textit{AND}-decomposition also satisfies both \textit{entailment} and \textit{consistency}. Following the bottom-up approach, we now proceed to evaluate the cumulative satisfaction condition of goal \textit{G} using equation \ref{eq:ORrec} as follows - 
\begin{align*}
CE(G) &= ORrec(G,G_1,G_2)\\
&= \{rec(G,G_1), rec(G,G_2)\} \\
rec(G,G_1) &= \{\{p,\neg q,w\} \cup \{s\}\} = \{p,\neg q,w,s\},\\ &\hspace{2cm} \text{and } [CE(G_1)\cap \neg IE(G)=\emptyset]\\
rec(G,G_2) &= \{\{p,\neg q\} \cup \emptyset\} = \{p,\neg q\},\\ &\hspace{2cm} \text{and } [CE(G_2)\cap \neg IE(G)=\{\neg w\}]\\
\\
\therefore CE(G) &= \{\{p,\neg q,w,s\},\{p,\neg q\}\}
\end{align*}
The cumulative satisfaction conditions of goal \textit{G} has been labelled in the figure. Since $IE(G)\subseteq\{p,\neg q,w,s\}$, we can say that there is at least one strategy that fulfils \textit{entailment}. However, since $CE(G_2)\cap \neg IE(G)$ is not null, we conclude that \textit{consistency} is not ensured in the second strategy and the conflicting satisfaction conditions are the members of the set $CE(G_2)\cap \neg IE(G)$, i.e., \{$w$\}.

\begin{figure}[h]
	\centering
	\includegraphics[width=0.9\textwidth, page=14]{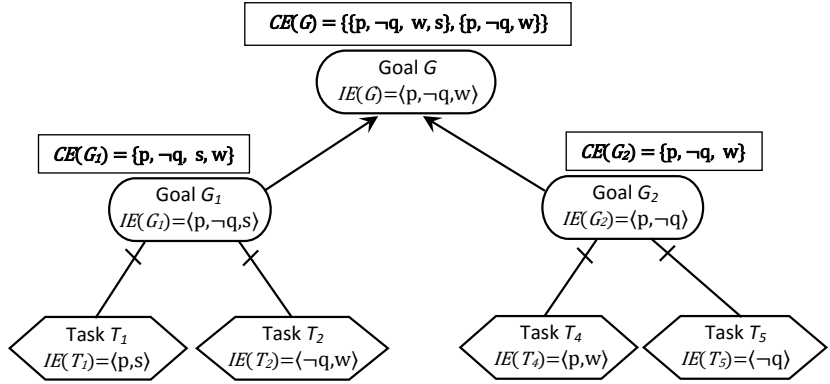}
	\caption{An example showing how \textit{entailment} and \textit{consistency} are both satisfied.}
	\label{fig:ec11}
\end{figure}

\paragraph*{Case 3: Both Entailment and Consistency are satisfied}
Consider the i$^*$ model shown in Figure \ref{fig:ec11}. This figure is exactly similar to figure \ref{fig:ec10} with one minor change. The immediate satisfaction condition of task $T_4$ is changed from $\{p,\neg w\}$ to $\{p, w\}$.

This is the best possible outcome that requirement analysts and the client would like to achieve in the requirements phase. The reconciliation of satisfaction conditions is done in the same way as shown in the previous two examples. Assuming that the reader has understood the working principle, we skip the cumulative satisfaction condition evaluation process. However, there are three interesting observations in this example that needs to be highlighted.
\begin{enumerate}
	\item Unlike the previous example shown in figure \ref{fig:ec10}, $CE(G_2) \cap \neg IE(G) = \emptyset$. This implies that the inconsistency issue existing in the previous example, does not persist in this scenario. Also, since $IE(G) \subseteq CE(G)$ , \textit{entailment} is satisfied.
	\item This example also justifies our bottom-up approach. None of the subgoals $G_1$ or $G_2$ have $w\in IE(G_1)$ or $w\in IE(G_2)$. However, satisfaction reconciliation ensures that the effect $w$ gets propagated from tasks $T_2$ and $T_4$  upwards such that all members of the cumulative satisfaction condition set $CE(G)$ satisfy the immediate satisfaction conditions $IE(G)$.
	\item This is the kind of situation where \textit{minimality} can play some role for system designers. Both members of the set $CE(G)$ satisfy the \textit{entailment} and \textit{consistency} conditions. In such a situation, designers may choose a particular strategy that produces a minimal set of additional satisfaction conditions. In this example, $\{p,\neg q,w\}$ is a more minimal solution for satisfying goal $G$ as compared to $\{p,\neg q,w, s\}$ as the latter produces an additional satisfaction condition of $\{s\}$.
\end{enumerate}

\paragraph*{Case 4: Neither Entailment nor Consistency are satisfied}

From the previous examples, one can easily visualize a scenario where neither \textit{entailment} nor \textit{consistency} is satisfied. This scenario is not at all desirable from the client's as well as designer's perspective. Requirement engineers may have to revisit the client and perform refinements of the previously elicited requirement specifications. The analysts can also help the client by highlighting erroneous and conflicting requirement specifications.

\subsection*{\textbf{Satisfaction Reconciliation Algorithm (\texttt{SRA})}}
\noindent \texttt{Input:} An i$^*$ (SR) model whose model artefacts have been annotated with immediate satisfaction conditions
\\

\noindent \texttt{Output:} Annotation of the model elements with cumulative satisfaction conditions derived from the satisfaction reconciliation process
\\

\noindent \texttt{Algorithm\_SRA:}
\begin{list}{\textit{Step-\arabic{qcounter}}:~}{\usecounter{qcounter}}
	\item Start.
	\item Identify a $ORGMod$ using the $ORGMod$ Extraction Algorithm.
	\item For each $ORGMod$ repeat the following steps. 
	\begin{list}{\textit{\alph{acounter}})~}{\usecounter{acounter}}
		\item Begin at the leaf level goals and check the existence of any dependencies. Evaluate the cumulative satisfaction condition $CE(L_a)$ of each leaf-level goal $L_a$ using Eqn.\ref{eq:DEPrec_mult}.
		
		\item Go to the previous level and check if the goals in this level undergo an AND-decomposition or an OR-decomposition. Depending on the type of merge operation required, evaluate the respective cumulative satisfaction conditions using Eqns.\ref{eq:ANDrec} or \ref{eq:ORrec}, respectively.
		
		\item Repeat the previous two steps till we reach the root of the extracted $ORGMod$.
	\end{list}	
	\item Check if the goal satisfaction condition set $IE()$ and the cumulative satisfaction condition set $CE()$ of the root satisfies \textit{entailment} and \textit{consistency}. If not, then raise a flag to the requirement analysts.
	\item Repeat Steps 3-4 for all possible $ORGMod$s.
	\item Stop.
\end{list}

\section{Resolving conflicts using model refactoring}
\label{sec:ResCon}
The \texttt{SRA} algorithm helps in identifying \textit{entailment} and \textit{consistency} issues during the bottom-up satisfaction reconciliation process. However, the AFSCR framework does not merely identify these issues. It also makes an attempt to resolve these issues by refactoring the given i$^*$ model. Requirement analysts are provided with possible solutions and necessary changes that need to be incorporated into the requirements model in order to satisfy \textit{entailment} and \textit{consistency}. In the following sections we first illustrate how we attempt to resolve these issues using test cases. We then propose a formal algorithm for doing the same.

\subsection{\textit{Entailment} issues}
The \texttt{SRA} algorithm raises an \textit{entailment }issue at any point in the satisfaction reconciliation process when the  satisfaction conditions $IE(M)$ of a goal $M$ are not achieved or satisfied by the sub-model rooted at $M$. The cumulative satisfaction condition of the sub-model are captured in $CE(M)$. Depending on whether $M$ undergoes an AND-decomposition or an OR-decomposition, $CE(M)$ contains only one member or multiple members, respectively. The cumulative satisfaction condition for AND-decompositions is one single set of satisfaction conditions obtained using Eqn.\ref{eq:ANDrec} defined in Section \ref{ANDmerge}. On the other hand, the cumulative satisfaction condition set for OR-decompositions contains as many members as the number of alternate strategies captured in the OR-decomposition. Each member is again a set of satisfaction conditions reconciled over that particular $ORGMod$, obtained using Eqn.\ref{eq:ORrec} defined in Section \ref{ORmerge}.

It is easy to raise an \textit{entailment} issue for AND-decompositions as we only need to check if $IE(M)\nsubseteq CE(M)$. However, for OR-decompositions, we need to check this condition for each individual member of the $CE(M)$ set. In general, we can formally define the condition for raising an \textit{entailment} issue as:
\begin{equation}
\exists CE_i\in CE(M), \text{ s.t. } IE(M)\nsubseteq CE_i \nonumber
\end{equation}
Once an \textit{entailment} issue is flagged by the \texttt{SRA} algorithm, we proceed to derive two data sets for resolving the issue - \texttt{deficiency-list}s and \texttt{availability-tuple}s. The \texttt{deficiency-list} $\mathbb{D}$ is used to identify all those immediate satisfaction conditions in $IE(M)$ that are not present in $CE(M)$. This is obtained individually for all members $CE_i$ of $CE(M)$. The set is evaluated as follows:
\begin{equation}
\mathbb{D}=\{IE(M)\setminus CE_i | \forall CE_i\in CE(M)\} 
\label{eq:def_lst}
\end{equation}
Once we obtain the \texttt{deficiency-list} $\mathbb{D}$ we proceed to explore whether these satisfaction conditions are fulfilled or achieved by artefacts that lie in other solution paths. We consider a one-to-one mapping, called the \textit{Availability Function}, that maps each member $d_i\in\mathbb{D}$ to a tuple of integers $\langle n_1,n_2,\cdots,n_k\rangle$. The \textit{Availability} mapping tries to capture the information whether any particular satisfaction condition $d_{ij}\in d_i$ can be fulfilled along other solution paths. It is defined as follows:
\begin{equation}
\mathbb{A}\colon\mathbb{D}\rightarrow\mathbb{N}^\Bbbk \nonumber
\end{equation}
such that $\mathbb{A}(d_i)$=$\langle n_1,n_2,\cdots, n_k\rangle$ where $k=|d_i|$ and $\forall d_{ij}\in d_i$,
\begin{equation}
n_j = \begin{cases}
r, \hspace{0.7cm}\text{if $\exists CE_r\in$CE(M) s.t. $d_{ij}\in CE_r$}.\\
0, \hspace{0.7cm}\text{otherwise}.
\end{cases} 
\label{eq:av_tup}
\end{equation}
Each such tuple corresponding to a \texttt{deficiency-list} is called an \texttt{availability tuple}. The set of all \texttt{availability-tuples} forms the range of the \textit{Availability Function}. In the next two sections, we demonstrate the model refactoring strategies that can be used to resolve \textit{entailment} issues for OR-decompositions and AND-decompositions.

\subsubsection{Entailment Resolution for OR-decompositions}
\begin{figure}[!h]
	\centering
	\includegraphics[width=0.5\textwidth, page=16]{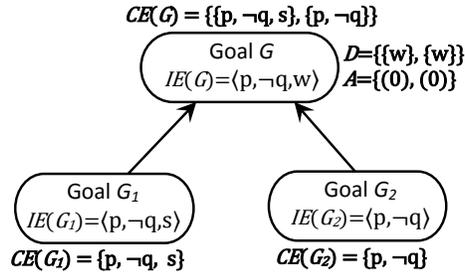}
	\caption{A sample i$^*$ model showing failure of \textit{entailment} at goal $G$ that undergoes OR-decomposition.}
	\label{fig:ent_ME}
\end{figure}

Consider the i$^*$ model shown in Figure \ref{fig:ent_ME}. There are two alternate means $G_1$ and $G_2$ for fulfilling the high-level goal $G$. Neither of the members in $CE(G)$ contains all the goal satisfaction conditions in $IE(G)$. We proceed to resolve this \textit{entailment} issue by first listing the \texttt{deficiency-list} for each path and then evaluating the \texttt{availability-tuple} for each path.
\\

\noindent The \texttt{deficiency-list} for goal $G_1$ is given by:
\begin{equation}
d_1 = IE(G)\setminus ce_1 = \{p,\neg q,w\}\setminus\{p,\neg q,s\} = \{w\} \nonumber
\end{equation}
The \texttt{availability-tuple} for goal $G_1$ is given by $\mathbb{A}(d_1)=(0)$ since $w$ is not contained in $ce_2$. Similarly, the \texttt{deficiency-list} and \texttt{availability-tuple} for goal $G_2$ is obtained as $d_2=\{w\}$ and $\mathbb{A}(d_2)=(0)$.

\begin{figure}[h]
	\centering
	\includegraphics[width=0.9\textwidth, page=17]{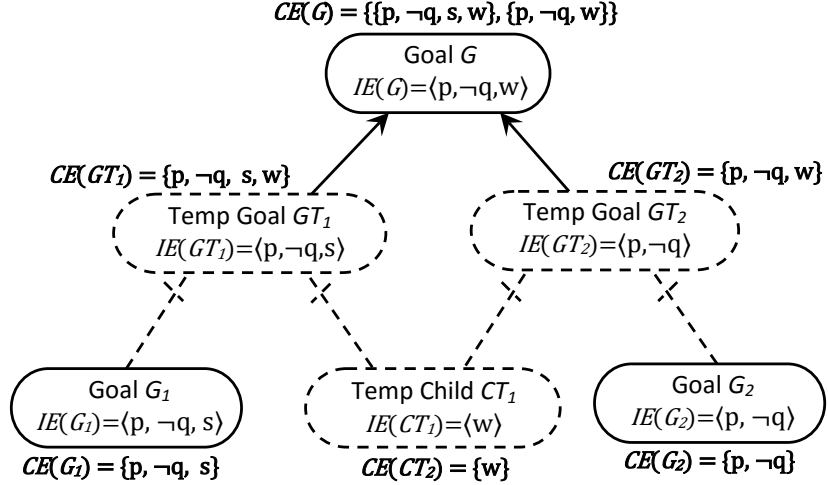}
	\caption{Temporary high-level goals $GT_1$ and $GT_2$ are used to merge goals $G_1$ and $G_2$ with the temporary goal $CT_1$.}
	\label{fig:ent_ME_res}
\end{figure}

This implies that the satisfaction condition\{$w$\} is not derived from either of the strategies. The intuition behind providing a solution to the requirement analysts is that \textquotedblleft\textit{we need to incorporate a goal, say $G^\prime$, which brings about the state of affairs \textquoteleft w\textquoteright on the world in which the actor resides}\textquotedblright. Thus, we introduce a temporary goal $CT_1$ with goal satisfaction condition $IE(CT_1)=\{w\}$ and merge it with goals $G_1$ and $G_2$ to achieve the satisfaction condition \textquoteleft$w$\textquoteright in the cumulative satisfaction condition of $G$. The solution is shown in Figure \ref{fig:ent_ME_res}.

\subsubsection{Entailment Resolution for AND-decompositions}
\begin{figure}[t]
	\centering
	\includegraphics[width=0.5\textwidth, page=28]{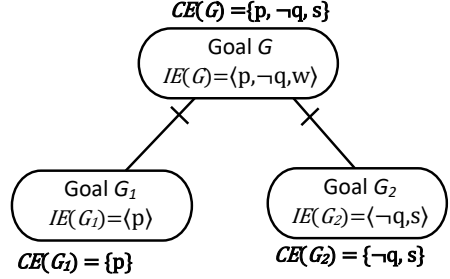}
	\caption{A sample i$^*$ model showing failure of \textit{entailment} at goal $G$ that undergoes AND-decomposition.}
	\label{fig:ent_TD}
\end{figure}

For AND-decompositions, the solution is not so complex. Consider the \textit{entailment} issue being addressed in Figure \ref{fig:ent_TD}. Since goal $G$ undergoes an AND-decomposition, it's cumulative satisfaction condition set $CE(G)$ contains only one member which is the set of satisfaction conditions derived from all the individual AND-decomposition links.

We proceed to evaluate the \texttt{deficiency-list} as follows:
\begin{equation}
d = IE(G)\setminus CE(G) = \{p,\neg q,w\}\setminus\{p,\neg q,s\} = \{w\} \nonumber
\end{equation}
The \texttt{availability-tuple} for goal $G$ is given by $\mathbb{A}(d)=(0,0)$. In fact, for AND-decompositions, the \texttt{deficiency-list} will always contain only one set member and it's \texttt{availability-tuple} will always be of the form (0,0,$\cdots$,0) depending on the number of satisfaction conditions in the \texttt{deficiency-list} set. 

\begin{figure}[t]
	\centering
	\includegraphics[width=0.9\textwidth, page=29]{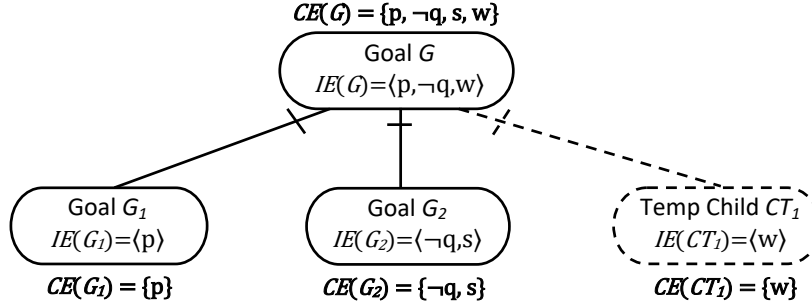}
	\caption{Temporary goal $CT_1$ is added to merge it with goals $G_1$ and $G_2$ for resolving the entailment conflict.}
	\label{fig:ent_TD_res}
\end{figure}

The reason is quite intuitive. The very semantics of an AND-reconciliation necessitate that all distinct satisfaction conditions appearing in all individual paths be reconciled in one set. This, in turn, makes the solution very simple - \textquotedblleft\textit{we need to incorporate a goal, say $G^\prime$, which brings about these unaccounted state of affairs (as obtained in the \texttt{deficiency-list}) on the world in which the actor resides}\textquotedblright. We introduce a temporary goal $CT_1$ with goal satisfaction condition $IE(CT_1)=\{w\}$ and merge it with goals $G_1$ and $G_2$ to incorporate these satisfaction conditions in the cumulative satisfaction condition of $G$. The solution is shown in Figure \ref{fig:ent_TD_res}.

\subsection*{\textbf{Entailment Resolution Algorithm (\texttt{ERA})}}
\noindent \texttt{Input:} Identify a goal $M$ as a point of \textit{entailment} failure if the following condition is satisfied:
\begin{equation}
\exists CE_i\in CE(M), \text{ s.t. } IE(M)\nsubseteq CE_i \nonumber
\end{equation}

\noindent \texttt{Output:} A possible solution for entailment resolution using model refactoring
\\

\noindent \texttt{Algorithm\_ERA:}
\begin{list}{\textit{Step-\arabic{qcounter}}:~}{\usecounter{qcounter}}
	\item Start.
	\item Evaluate the \textit{Deficiency} set $\mathbb{D}$ using Eqn.\ref{eq:def_lst}.
	
	\item Define the \textit{Availability} mapping function $\mathbb{A}$  according to Eqn.\ref{eq:av_tup}.
	
	\item If the point of failure $M$ undergoes OR-Decomposition and produces nodes $M_1,M_2,\cdots,M_p$, then propagate the respective \texttt{deficiency-list} and the corresponding \texttt{availability-tuple} along each of the $p$ paths. For each path, do the following:
	\begin{list}{\textit{\alph{acounter}})~}{\usecounter{acounter}}
		\item At the next level, create a temporary high level-goal $GT_i$ having the same goal satisfaction conditions as $M_i$, i.e., $IE(GT_i)=IE(M_i)$.
		\item Create an AND-decomposition of $GT_i$ with its leftmost child being $M_i$.
		\item Add another child $CT_i$ to $GT_i$ whose goal satisfaction conditions are obtained by concatenating those members in the \texttt{deficiency-list} $d_i$ whose corresponding \texttt{availability-tuple} values $n_j$ are zero (0).
		\item For every other member $d_{ij}$ in the \texttt{deficiency-list} that has a non-zero \texttt{availability-tuple} value $n_j$, set up an AND-decomposition link (if it does not exist already) between $GT_i$ and the goal $M^\prime_i$ residing in path $n_j$ such that $d_{ij}\in IE(M^\prime_i)$.
	\end{list}
	
	\item If the point of failure $M$ undergoes AND-decomposition and produces nodes $M_1,M_2,\cdots,M_q$, then add another temporary child $CT_i$ under $M$ and set up an AND-decomposition link between $M$ and $CT_i$. Annotate $CT_i$ as $CE(CT_i) = IE(CT_i)=\mathbb{D}$.
	
	\item Repeat Steps 2-5 for all goals, in a bottom-up manner, during the satisfaction reconciliation procedure.
	
	\item Stop.
\end{list}

\subsection{\textit{Consistency} issues}
Consistency is defined as a condition where the cumulative satisfaction condition of any goal $M$ does not contain mutually conflicting immediate satisfaction conditions. Otherwise, the goal is said to be inconsistent. Inconsistency of goals during the satisfaction reconciliation process can be classified into two different types - \textit{hierarchic inconsistency} and \textit{sibling inconsistency}.

\subsubsection{Hierarchic Inconsistency}

\begin{figure}[h]
	\centering
	\includegraphics[width=0.5\textwidth, page=20]{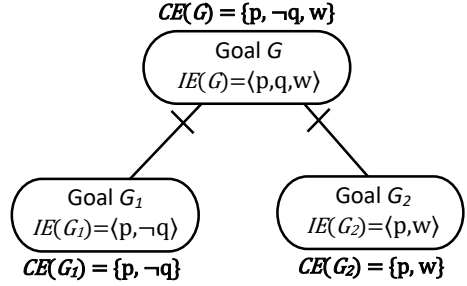}
	\caption{\textit{Hierarchic} inconsistency at goal $G$ arising out of the goal satisfaction condition $q$ of $G$ and the cumulative satisfaction condition $\neg q$ of goal $G_1$.}
	\label{fig:hier_incons}
\end{figure}

\textit{Hierarchic} inconsistency occurs when some immediate satisfaction condition(s) of a parent goal is in conflict with some cumulative satisfaction condition of a child goal. This type of inconsistency can occur for both AND-decompositions and OR-decompositions. In case of OR-decompositions, we need to be worried with only those alternate means that are inconsistent. The system does not fail as long as there exists one alternative that is consistent with respect to satisfaction conditions. For AND-decompositions, the consequences are much more critical and can result in an inconsistent system. Figure \ref{fig:hier_incons} illustrates an AND-decomposition that results in hierarchic inconsistency.

The \texttt{SRA} algorithm takes care of \textit{hierarchic} inconsistency in the satisfaction reconciliation process itself. Equation \ref{eq:rec} of Theorem \ref{TH:FEA} removes all those cumulative satisfaction conditions of the child goal that are inconsistent with goal satisfaction conditions of the parent goal while evaluating the cumulative satisfaction conditions of the parent goal.

\begin{figure}[h]
	\centering
	\includegraphics[width=0.7\textwidth, page=21]{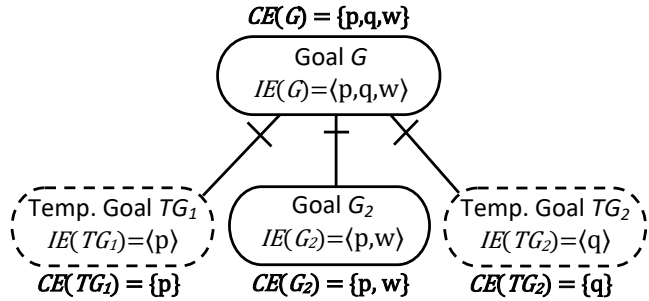}
	\caption{Eliminating inconsistencies in the satisfaction reconciliation process governed by Equation \ref{eq:rec} in Theorem \ref{TH:FEA}.}
	\label{fig:hier_cons}
\end{figure}

Figure \ref{fig:hier_cons} shows how hierarchic inconsistency is tackled in the \texttt{SRA} algorithm. We replace the child goal $G_1$ with a temporary goal $TG_1$ whose immediate satisfaction set does not contain the satisfaction conditions that conflict with the parent goal $G$. Now, the problem basically reduces to that of \textit{Entailment Resolution} and the \texttt{ERA} algorithm can be used to resolve it.

\subsubsection{Sibling Inconsistency}
\begin{figure}[h]
	\centering
	\includegraphics[width=0.5\textwidth, page=22]{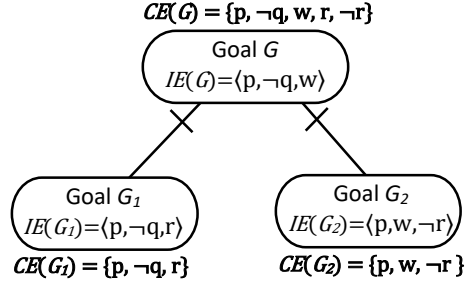}
	\caption{\textit{Sibling} inconsistency at goal $G$ arising out of the goal satisfaction condition $r$ of goal $G_1$ and the goal satisfaction condition $\neg r$ of goal $G_2$.}
	\label{fig:sibl_incons}
\end{figure}

The other type of inconsistency is \textit{sibling} inconsistency which arises during satisfaction reconciliation of mutually conflicting satisfaction conditions from child nodes of the same parent node. Figure \ref{fig:sibl_incons} illustrates one such scenario where the cumulative satisfaction conditions of goal $G$ contain both $r$ and $\neg r$, reconciled from child goals $G_1$ and $G_2$, respectively. 

\begin{figure}[htp]
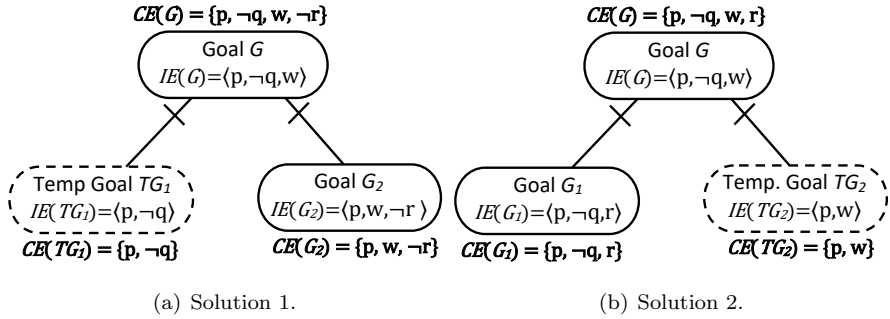

	\begin{center}
		\subfigure[Solution 1.]{\label{fig:sibl_cons1}\includegraphics[width=0.48\textwidth, page=23]{figures.pdf}}
		\subfigure[Solution 2.]{\label{fig:sibl_cons2}\includegraphics[width=0.48\textwidth, page=24]{figures.pdf}}
	\end{center}
	\caption{\texttt{Solution 1}: eliminates the satisfaction condition $r$ of goal $G_1$. \texttt{Solution 2}: eliminates the satisfaction condition $\neg r$ of goal $G_2$.}
	\label{fig:sibl_cons}
\end{figure}

Unlike resolution of hierarchic inconsistencies, resolving sibling inconsistencies do not result in entailment issues. This implies that removing any one of the conflicting satisfaction conditions from the child goals is sufficient to resolve this type of inconsistency. With respect to the scenario shown in Figure \ref{fig:sibl_incons}, we can highlight to the analysts that either effect $r$ of goal $G_1$ or effect $\neg r$ of goal $G_2$, needs to be dealt with. These two solutions are shown in Figure \ref{fig:sibl_cons}.

The requirement analysts can then decide which particular solution best suits the requirements of the enterprise depending on the consequences and significance of the satisfaction conditions $r$ and $\neg r$.

\subsection*{\textbf{Consistency Resolution Algorithm (\texttt{CRA})}}
\noindent \texttt{Input:} Identify a goal $M$ as a point of \textit{consistency} failure.
\\

\noindent \texttt{Output:} A possible solution for consistency resolution using model refactoring
\\

\noindent \texttt{Algorithm\_CRA:}
\begin{list}{\textit{Step-\arabic{qcounter}}:~}{\usecounter{qcounter}}
	\item Start.
	
	\item Identify the type of inconsistency as \textit{hierarchic} if - 
	\begin{equation}
	\exists ce_i\in CE(M), ie_j\in IE(M) \text{ s.t. } ce_i = \neg ie_j \nonumber
	\end{equation}
	
	\begin{list}{\textit{\alph{acounter}})~}{\usecounter{acounter}}
		\item Identify the child $M_k$ which contributes the satisfaction condition $ce_i$ in $CE(M)$, i.e., $ce_i\in CE(M_k)$.		
		\item Remove the satisfaction condition $ce_i$ from the cumulative satisfaction condition set as well as its goal satisfaction condition set, i.e.,
		\begin{align*}
		CE^\prime(M_k) &= CE(M_k)\setminus ce_i\\
		IE^\prime(M_k) &= IE(M_k)\setminus ce_i
		\end{align*}
		\item The i$^*$ model is now consistent but the goal satisfaction condition $ie_j$ of $M$ does not appear in its cumulative satisfaction condition set. This can be resolved by calling the \texttt{ERA} algorithm.
	\end{list}
	
	\item Identify the type of inconsistency as \textit{sibling} if - 
	\begin{align*}
	\exists ce_i,ce_j\in CE(M), ce_i,ce_j\notin IE(M) \text{ s.t. } (ce_i=\neg ce_j)\nonumber
	\end{align*}
	
	\begin{list}{\textit{\alph{acounter}})~}{\usecounter{acounter}}
		\item Identify the siblings $M_{k1}$ and $M_{k2}$ which contribute the satisfaction conditions $ce_i$ and $ce_j$ in $CE(M)$, respectively. That is, 
		\begin{align*}
		ce_i\in CE(M_{k1}) \text{ and } ce_j\in CE(M_{k2}).		
		\end{align*}
		
		\item Remove the satisfaction condition $ce_i$ from the cumulative and immediate satisfaction condition sets of $M_{k1}$, i.e.,
		\begin{align*}
		CE^\prime(M_{k1}) &= CE(M_{k1})\setminus ce_i\\
		IE^\prime(M_{k1}) &= IE(M_{k1})\setminus ce_i
		\end{align*}
		
		\item Remove the satisfaction condition $ce_j$ from the cumulative and immediate satisfaction condition sets of $M_{k2}$, i.e.,
		\begin{align*}
		CE^\prime(M_{k2}) &= CE(M_{k2})\setminus ce_j\\
		IE^\prime(M_{k2}) &= IE(M_{k2})\setminus ce_j
		\end{align*}
		
		\item Present both the above alternatives to the requirement analysts as consistent solutions.
	\end{list}
	
	\item Repeat Steps 2-3 for all goals, in a bottom-up manner, during the satisfaction reconciliation procedure.
	
	\item Stop.
\end{list}

The AFSCR framework has been proposed as a support tool for requirement analysts and enterprise architects to perform satisfaction analysis of goal models. The very notion of the framework is to guide requirement analysts so that they can identify and resolve satisfaction conflicts which may be captured in goal models. The framework is a semi-automated framework that identifies satisfaction conflicts and proposes suggestive conflict resolutions to the requirement analysts. The conflict resolutions (both entailment and consistency) that are performed by the AFSCR framework are minimal in nature, i.e., the authors propose a workaround that takes care of the conflict and yet the new goal model deviates minimally from the previous version of the goal model. This is only a proposal for the enterprise architects and it is up to them to accept the solution or go for a completely new goal model based on their own interpretations of the conflict that is identified. The temporary tasks which may be created to resolve a conflict only suggest requirement analysts to rethink the design of the goal model.

\section{Use Case: Healthcare}
\label{sec:UseCase}
\begin{figure}[t]
	\centering
	\includegraphics[width=0.9\textwidth, page=26]{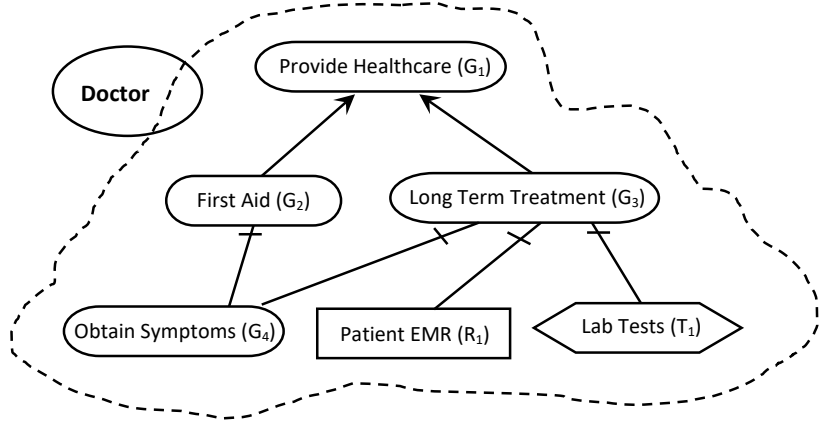}
	\caption{Goal model of an existing healthcare enterprise.}
	\label{fig:ex2_HC}
\end{figure}

In this section, we take a real life use case of a healthcare enterprise. We will first demonstrate how a goal model can be annotated in real world business settings. We will also demonstrate how we can apply the AFSCR framework for identification of conflicts and their resolution in an evolving environment with changing business demands. Let us consider the healthcare example shown in figure \ref{fig:ex_RHC} and redrawn in figure \ref{fig:ex2_HC} with goal, task and resource labels.

The goal model in figure \ref{fig:ex2_HC} can be annotated as follows:
\begin{itemize}
	\item \texttt{IE(G$_1$)} = \{\textit{Received\_Patient, Provided\_Relief}\}
	\item \texttt{IE(G$_2$)} = \{\textit{Emergency\_Treatment\_Provided}\}
	\item \texttt{IE(G$_3$)} = \{\textit{Normal\_Treatment\_Provided}\}
	\item \texttt{IE(G$_4$)} = \{\textit{\{Received\_Text\}, \{Received\_Voice\}}\}
	\item \texttt{IE(R$_1$)} = \{\textit{PreExisting\_Disease\_Searched, Allergies\_Checked}\}
	\item \texttt{IE(T$_1$)} = \{\textit{\{\{Sample\_Taken\},\{Performed\_Procedure\}\}, Test\_Result\_Known}\}
\end{itemize}

Annotation of individual goals, tasks, and resources within the goal model with immediate satisfaction conditions is the only phase of the AFSCR framework that requires human intervention. These annotations are context-free and the requirement analyst lists them for each goal model artifact in a stand-alone perspective. Thus, the immediate satisfaction conditions need to be associated with a knowledge base of semantic rules that correlate these immediate satisfaction condition. Thus, for the above set of immediate satisfaction conditions, we have the following correlation rules:
\begin{list}{\textbf{KB\arabic{acounter}:}~}{\usecounter{acounter}}
	\item \textit{Emergency\_Treatment\_Provided} $\rightarrow$ \textit{Received\_Text} $\vee$ \textit{Received\_Voice}.
	\item \textit{Normal\_Treatment\_Provided} $\rightarrow$ (\textit{Received\_Text} $\vee$ \textit{Received\_Voice}) $\wedge$ \textit{PreExisting\_Disease\_Searched} $\wedge$ \textit{Test\_Result\_Known}.
	\item \textit{Received\_Patient} $\rightarrow$ \textit{Received\_Text} $\vee$ \textit{Received\_Voice}
	\item \textit{Provided\_Relief} $\rightarrow$ \textit{Emergency\_Treatment\_Provided} $\vee$ \textit{Normal\_Treatment-\_Provided}.
\end{list}

Using the AFSCR framework rules and the knowledge base rules, we can compute the cumulative satisfaction conditions for each goal model artifact as follows:
\begin{itemize}
	\item \texttt{CE(G$_4$)} = \{\textit{\{Received\_Text\}, \{Received\_Voice\}}\}
	\item \texttt{CE(R$_1$)} = \{\textit{PreExisting\_Disease\_Searched, Allergies\_Checked}\}
	\item \texttt{CE(T$_1$)} = \{\textit{\{\{Sample\_Taken\}, \{Performed\_Procedure\}\}, Test\_Result\_Known}\}
	\item \texttt{CE(G$_2$)} = \{\textit{Emergency\_Treatment\_Provided, \{\{Received\_Text\}, \{Received\_Voice\}\}}\}\\
	Applying \textbf{KB1}, the CE set for $G_2$ reduces to:\\
	\texttt{CE(G$_2$)} = \{\textit{\{Received\_Text\}, \{Received\_Voice\}}\}
	\item \texttt{CE(G$_3$)} = \{\textit{Normal\_Treatment\_Provided, \{\{Received\_Text\}, \{Received\_Voice\}\}, PreExisting\_Disease\_Searched, Allergies\_Checked, \{\{Sample\_Taken\}, \{Performed\_Procedure\}\}, Test\_Result\_Known}\}\\
	Applying \textbf{KB2}, the CE set for $G_3$ reduces to:\\
	\texttt{CE(G$_3$)} = \{\textit{\{\{Received\_Text\}, \{Received\_Voice\}\}, PreExisting\_Disease\_Searched, Allergies\_Checked, \{\{Sample\_Taken\}, \{Performed\_Procedure\}\}, Test\_Result\_Known}\}
	\item \texttt{CE(G$_1$)} = \{\textit{\{Received\_Patient, Provided\_Relief, \{\{Received\_Text\}, \{Received\_Voice\}\}\}, \{Received\_Patient, Provided\_Relief, \{\{Received\_Text\}, \{Received\_Voice\}\}, PreExisting\_Disease\_Searched, Allergies\_Checked, \{\{Sample\_Taken\}, \{Performed\_Procedure\}\}, Test\_Result\_Known}\}\}\\
	Applying rules \textbf{KB3} and \textbf{KB4}, the CE set of $G_1$ reduces to:\\
	\texttt{CE(G$_1$)} = \{\textit{\{\{Received\_Text\}, \{Received\_Voice\}\}, \{\{\{Received\_Text\}, \{Received\_Voice\}\}, PreExisting\_Disease\_Searched, Allergies\_Checked, \{\{Sample\_Taken\}, \{Performed\_Procedure\}\}, Test\_Result\_Known}\}\}
\end{itemize}

Based on this set of computed cumulative satisfaction conditions, we can now check that the goal model is free from all types of conflicts and consistent with the business environment. Now let us suppose a couple of changes in the business environment setting:

\begin{list}{\texttt{Change-\arabic{acounter}:}~}{\usecounter{acounter}}
	\item The healthcare enterprise passes a regulation that \textit{Long Term Treatment} cannot be provided without consulting a specialist.
	\item It must be ensured that patient is not allergic to any chemicals before performing a test. This is to prevent situations such as an MRI scan (with contrast) becomes the reason of death for a patient who is allergic to contrast fluids like iodine.
\end{list}

These changes in the business environment can be reflected in the goal model of figure \ref{fig:ex2_HC} by updating the immediate satisfaction conditions of goal $G_3$  (for \texttt{Change-1}) and task $T_1$ (for \texttt{Change-2}). The modified immediate satisfaction sets of these two goal model artifacts become as follows:
\begin{itemize}
	\item  Modified \textbf{KB2:} \textit{Normal\_Treatment\_Provided} $\rightarrow$ (\textit{Received\_Text} $\vee$ \textit{Received\_Voice}) $\wedge$ \textit{PreExisting\_Disease\_Searched} $\wedge$ \textit{Test\_Result\_Known} $\wedge$ \textit{Consulted\_Specialist}.\\
	\texttt{IE$^\prime$(G$_3$)} = \{\textit{Normal\_Treatment\_Provided}\} $\equiv$ \{\textit{\{\{Received\_Text\}, \{Received\_Voice\}\}, PreExisting\_Disease\_Searched, Test\_Result\_Known, Consulted\_Specialist}\}.
	\item \texttt{IE$^\prime$(T$_1$)} = \{\textit{Allergies\_Checked, \{\{Sample\_Taken\},\{Performed\_Procedure\}\}, Test\_Result\_Known}\}
\end{itemize}

\begin{figure}[t]
	\centering
	\includegraphics[width=0.9\textwidth, page=27]{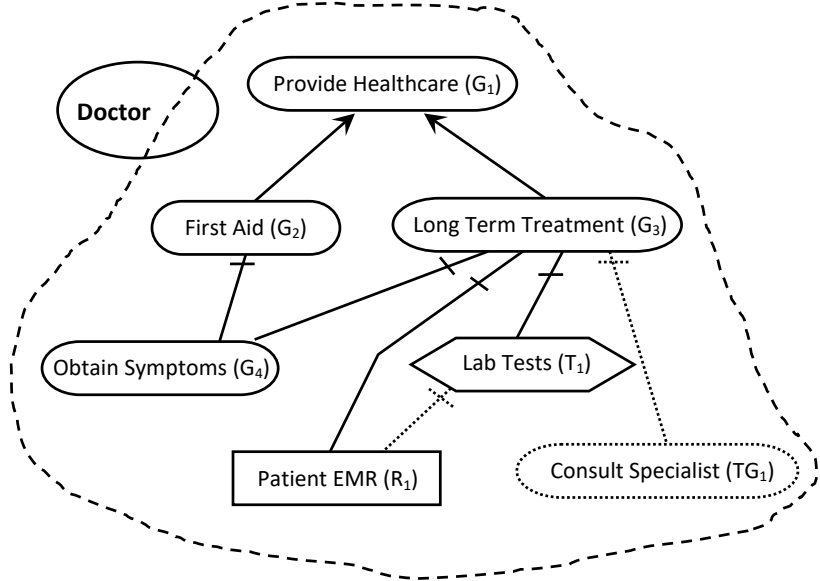}
	\caption{Modified goal model incorporating the two business environment changes \texttt{Change-1} and \texttt{Change-2}.}
	\label{fig:ex3_HC}
\end{figure}

If we now use the AFSCR framework to compare the cumulative satisfaction conditions (already computed for the previous business setting) with the modified immediate satisfaction conditions (in the current setting), we will notice \textit{entailment} conflicts for both $G_3$ and $T_1$. The AFSCR framework handles these two conflicts with the Entailment Resolution Algorithm separately as follows:
\begin{list}{\texttt{ERA-\arabic{acounter}:}~}{\usecounter{acounter}}
	\item The entailment conflict for $G_3$ is due to the newly introduced satisfaction condition $\langle Consulted\_Specialist\rangle$. Since this condition is not fulfilled by any of the child nodes and $G_3$ undergoes AND-decomposition, we add another temporary goal called \textquotedblleft Consult Specialist\textquotedblright as a child of $G_3$. We label this goal $TG_1$.
	\item For the entailment conflict of $T_1$, the AFSCR framework finds that the newly added satisfaction condition $\langle Allergies\_Checked\rangle$ can be fulfilled by it's sibling $R_1$. So the AFSCR framework sets up a parent child link between $T_1$ and $R_1$.
\end{list}
The modified goal model that incorporates these changes is shown in figure \ref{fig:ex3_HC}.

\section{Analysis of the algorithms}
\label{sec:Analysis}
The AFSCR framework consists of three algorithms - Satisfaction Reconciliation, Entailment Resolution, and Consistency Resolution. Let us discuss the \textit{soundness}, \textit{completeness} and \textit{finiteness} properties of the three algorithms. 

The Satisfaction Reconciliation Algorithm (\texttt{SRA}) satisfies both the properties. The algorithm is \textit{sound} because it correctly performs the satisfaction reconciliation process whenever it is applied on an annotated goal model specification. The algorithm also correctly identifies all the entailment and consistency conflicts that exist. \texttt{SRA} is also \textit{complete} because it can perform the reconciliation and conflict-identification processes for every goal model specification whose model elements have been annotated with the immediate satisfaction conditions.

The Entailment Resolution Algorithm (\texttt{ERA}) and Conflict Resolution Algorithm (\texttt{CRA}) are both \textit{complete} because whenever a conflict is identified by \texttt{SRA}, depending on the type of conflict, \texttt{ERA} or \texttt{CRA} can resolve the conflict by making certain minimal refactoring of the goal model. However, the \textit{soundness} of the two algorithms cannot be ensured because resolving an entailment or consistency conflict does not guarantee that another conflict will not be introduced at some other node of the goal model. However, the AFSCR framework (as a whole) is \textit{sound} because \texttt{SRA}, \texttt{ERA}, and \texttt{CRA} are used iteratively by the framework to eliminate all conflicts using a top-down or bottom-up approach. Thus, for a given annotated goal model specification, we can ensure that the AFSCR framework can come up with a conflict-free goal model modification.

\noindent{\bf Proposition:} {\em Algorithms SRA, CRA and ERA terminate, given goal models with a finite number of goals as input, where the satisfaction condition of each goal is of finite length.} The AFSCR process involves a finite number of calls to the \texttt{SRA}, \texttt{ERA}, or \texttt{CRA} modules and, thus, is also guaranteed to terminate in finite time. However, if we cannot guarantee the finiteness of the goal model specification, then we cannot ensure \textit{finiteness} of the AFSCR process.

\section{Conclusion and Future Work}
\label{sec:Concl}
In this paper, we propose AFSCR as a semi-automated framework for satisfaction reconciliation of satisfaction conditions over annotated i$^*$ models. One of the major limitations of this work is that it depends on requirement engineers for the correct representation of immediate satisfaction conditions for a process. However, if this is assured then the satisfaction analysis and reconciliation process is fully automated. Another limitation of the work is that it has been proposed over the i$^*$  framework, using i$^*$ modelling  constructs. The choice of i$^*$ was conscious as it is one of the most powerful goal oriented requirements engineering framework. The AFSCR framework can be extended to other GORE frameworks like KAOS as well.

The precise nature of the formal assertions involved in the various satisfaction conditions of interest requires special attention. One way to represent these conditions is to restrict ourselves to propositional logic (we have done this in the examples in the paper for simplicity). However, the well-known limitations on the expressive power of propositional logic will always be a drawback (rules have to be written for every ground instance, for example, and assertions that hold for all objects – or at least one object – of interest cannot be easily written). Let us consider what might happen if we permit first-order logic assertions. Consider the assertion that the specialist knows the blood test results of a patient as a sub-goal in some part of the case study that we detail in Section \ref{sec:UseCase}. Let us use the predicate blood-test-results-known(S, P) to write this assertion. Here S is a variable that will be instantiated at runtime with the identifier of the specialist in question while P is a variable that will be instantiated with the identifier of the patient in question. The challenge now is to appropriately quantify these variables. Clearly, in this example, our intent is not to assert that all specialists known the blood-test results of all patients, hence the use of the universal quantifier is not warranted. It is also not our intent to assert that there exists at least one specialist who knows the test results of all patients, nor that there exists at least a specialist and a patient such that the specialist knows the blood-test results of that patient and so on. In other words, existentially quantifying these variables, or using combinations of existential and universal quantifiers do not serve our purpose. The most expressive means of writing this assertion is to write it as a first-order logic schema blood-test-results-known(S, P) where S and P remain unquantified. In a specific instance, when a specific specialist and patient are the objects of attention, we will obtain a ground instance of this schema, with the variables S and P instantiated with the appropriate identifiers. Ground instances of first-order schemas are effectively propositional assertions, which takes us back to the realm of propositional logic.

However, much of the reasoning that we will perform will be design-time reasoning, where instance-level information such as the identifiers of the specialist and patient in question will not be available. Here too the reasoning ultimately reduces to propositional reasoning. If there is only one specialist and one patient in question, the specific identifiers of these is immaterial to what we want to achieve. It is adequate to represent the goal or sub-goal assertion via the following proposition: patient-blood-test-results-known-to-specialist (if there were 2 specialists at play, we would create two versions of this proposition, one for each specialist, such as patient-blood-test-results-known-to-specialist1 and so on). In other words, first-order schemas are a convenient representation of these conditions to support ease of understand, but the reasoning ultimately reduces to propositional reasoning.

The AFSCR framework has been proposed as a solution for the goal model management problem. However, the framework lacks a proper tool or prototype that can verify and validate the proposed methodologies for satisfaction reconciliation. Goal model management is an optimization problem that is defined over an evolving business environment that is subject to changing business requirements. Goal models have to be robust and agile to incorporate such changes into the business processes. The goal model management problem tries to provide this flexibility and optimize the degree of change that is to be performed on a goal model. 

We are working on a tool interface for the AFSCR framework. This is the next immediate objective for verification and validation of the framework. The tool will allow us to test the AFSCR framework in changing business environments. The tool for implementing the framework proposes several challenges to the developers. The optimization problem tries to minimize the set of proposed changes for conflict management in case entailment and consistency conflicts are identified. Heuristics are being applied to resolve the optimization problem. The objective is to come up with a tool that makes minimal amount of changes to a given goal model state in order to resolve conflicts. Optimality will also ensure improvements in time complexity of the satisfaction reconciliation process - a critical factor for large and complex business processes.

\section*{Acknowledgement}
This work is a part of the Ph.D. work of Novarun Deb, who is a Research Fellow in the University of Calcutta under the Tata Consultancy Services (TCS) Research Scholar Program (RSP). We acknowledge the contribution of TCS Innovation Labs in funding this research. Part of this work was done by Novarun Deb at the Decision Systems Lab, University of Wollongong, during June-July 2014. We acknowledge the Technical Education Quality Improvement Programme (TEQIP), University of Calcutta, for organizing and sponsoring his visit to the university in Wollongong, Australia.

\bibliography{5000bib_mod}

\end{document}